\documentclass{aa}

\usepackage{graphicx,xspace}
\usepackage{natbib}
\usepackage{url}
\usepackage[varg]{txfonts}
\usepackage{threeparttable}
\usepackage{amsmath, amssymb}
\usepackage{graphicx,color}
\usepackage{microtype}
\usepackage{threeparttable}
\usepackage{booktabs, makecell}
\usepackage{subfig}
\usepackage{mwe}
\usepackage{subcaption}
\usepackage{lscape}

\def\ari{{Ariel V}\xspace}

\def\cxo{{Chandra}\xspace}
\def\ero{{eROSITA}\xspace}
\def\ede{{eROSITA\_DE}\xspace}
\def\eins{{Einstein}\xspace}
\def\exos{{EXOSAT}\xspace}
\def\gai{{Gaia}\xspace}
\def\heao{HEAO\xspace}

\def\ros{{ROSAT}\xspace}

\def\srgero{{SRG/eROSITA}\xspace}
\def\srg{{SRG}\xspace}
\def\swi{{Swift}/XRT\xspace}

\def\uhu{{UHURU}\xspace}
\def\xmmn{{XMM-Newton}\xspace}
\def\xspec{\texttt{{XSPEC}}\xspace}

\newcommand\fergs{\ensuremath{\mathrm{erg}\,\mathrm{cm}^{-2}\,\mathrm{s}^{-1}}\xspace}

\newcommand\fxo{${f_{\rm X}}/f_{\rm opt}$\xspace}

\newcommand\lx{\ensuremath{\mathrm{erg}\,\mathrm{s}^{-1}}\xspace}

\newcommand\tpar{{Table~\ref{t:parms}}\xspace}

\begin{document}

\title{A first systematic characterization of cataclysmic variables in \srgero surveys\thanks{Table~\ref{t:uniqdr1} is only available in electronic form at the CDS }} 
\author{A.D.~Schwope\inst{1}
\and K.~Knauff\inst{1}
\and J.~Kurpas\inst{1,2}
\and M.~Salvato\inst{3}
\and B.~Stelzer\inst{4}
\and L.~Stütz\inst{1}
\and D.~Tub\'in-Arenas\inst{1,2}
}
\institute{Leibniz-Institut f\"ur Astrophysik Potsdam (AIP), An der Sternwarte 16, 14482 Potsdam, Germany\\
\email{aschwope@aip.de}
\and 
Potsdam University, Institute for Physics and Astronomy, Karl-Liebknecht-Straße 24/25, 14476 Potsdam, Germany
\and
Max-Planck-Institut f\"ur extraterrestrische Physik, Gie{\ss}enbachstra{\ss}e, 85748 Garching, Germany
\and 
Institut für Astronomie \& Astrophysik, Eberhard Karls Universität Tübingen, Sand 1, 72076 Tübingen, Germany
}
\authorrunning{Schwope et al.}
\titlerunning{Systematic characterization of cataclysmic variables in \srgero surveys}
\date{\today}


\abstract{(abridged) We present an account of known cataclysmic variables (CVs) that were detected as X-ray sources in eROSITA X-ray surveys and have \textit{\gai} DR3 counterparts. We address standard CVs with main sequence donors and white dwarfs accreting via Roche-lobe overflow (RLOF) and related objects, the double degenerates (DDs), and the symbiotic stars (SySts). We discern between nonmagnetic (dwarf novae and nova-like objects) and magnetic CVs (polars and intermediate polars (IPs)). In the publically available eROSITA catalog from the recent DR1, typically 65\% of known cataloged and classified CVs are detected. This fraction rises to over 90\% if the stack of all \ero X-ray surveys (called S45 in this paper) is considered and the search volume is restricted to a radius of 500\,pc. We examine the various classes of CVs in various diagnostic diagrams relating X-ray and optical properties (luminosity, absolute magnitude, color, X-ray spectral hardness, and optical variability) and establish their average class properties. We derive spectral properties for the 22 brightest polars and confirm an increase in the ratio of soft to hard X-rays with increasing magnetic field in the accretion region. We report three new soft IPs and present a spectral analysis of all soft IPs. Their blackbody temperatures agree well with published values. The DDs represent the bluest and faintest subcategory but reach the same identification fraction as the standard CVs. The SySts are the most distant systems; only 20 (13\%) were detected as X-ray sources in S45, and 7 of those are first-time detections. We investigate their mean properties using an upper limit on the flux of the nondetected CVs. Their X-ray nondetection is indeed a  distance effect. }


\maketitle

\section{Introduction}

X-ray surveys have uncovered accreting compact white-dwarf binaries (CWDBs) since their conception. Here and in the following, we refer to accreting CWDBs as cataclysmic binaries (CBs) or cataclysmic variables (CVs). 
The two polars (magnetic CVs) AM Herculis and EF Eridani are among the brightest X-ray sources in the sky and were discovered with the \uhu and \ari satellites. The \heao\ observatories and \exos added further detections \citep{hertz+90, giommi+91}. A significant step forward was made with \ros, which uncovered many magnetic cataclysmic variables and allowed class properties to be studied \citep{beuermann+thomas93, beuermann+schwope94, schwope+02}  -- within
some limitations due to the still small sample sizes. 

\cite{gaensicke05} reviewed the various ways by which CVs have been found, namely through their variability properties, their optical color, their X-ray emission, and as a by-product of spectroscopic surveys (the SDSS in particular), and gave an account of the extent to which the selection mechanism affects fundamental population properties such as the orbital period distribution. Up to that point, none of the studied samples could be regarded as representative of the parent population. 
With \cxo and \xmmn, many detailed studies of individual objects were performed, but samples grew only slowly because of the small survey area of those missions and the low surface density of CVs. Meanwhile, \textit{\gai} has been shown to have a fundamental impact in many areas, including CV science. \cite{schwope18} derived an improved space density of nonmagnetic and magnetic CVs, \cite{pala+20} composed the first volume-limited sample within 150 pc, and \cite{abril+20} located the various subcategories in the 
color--magnitude diagram. 

Candidate CVs were found in large number by the Catalina Real-Time Transient Survey \citep[CRTS,][]{drake+14}. Over the more than two decades of its operation, the SDSS has observed more than 600 CVs of all types \citep{inight+23a}, serendipitously in many cases, and the ongoing SDSS-V project runs identification programmes to uncover CVs systematically \citep{inight+23b, schwope+24}. 

On the X-ray side, \ero on Spektrum-Roentgen-Gamma (\srg) is having a large impact on X-ray astronomy \citep{predehl+21, sunyaev+21}. The repeated sensitive surveys with an imaging telescope array in the energy range of $0.2 -10 $\,keV offer improved spatial and spectral resolution compared to its predecessor, the \ros all-sky survey (RASS). eROSITA allows large and largely unbiased samples of all types of CVs to be built for the first time. As shown in \cite{schwope+24}, past selection biases may be largely overcome by combining X-ray with UV selection. The occasion of the worldwide release of the first \ero all-sky survey \citep[eRASS1,][]{merloni+24} is a timely opportunity to review the power of \ero in building large samples of all types of CVs. A first glimpse of the anticipated results was provided by some serendipitous discoveries of new individual CVs in recent years, by the compilation of small samples, and by the first systematic study of CVs in the eFEDS survey \citep{munoz-giraldo+23, pelisoli+23, rodriguez+23, schwope+22b, schwope+22a, schwope+24}. Significantly enlarged CV samples as a result of the combination of \ero X-ray and UV-excess selection with \textit{\gai} parameters will provide sought-after constraints needed to address fundamental questions regarding close binary evolution and to derive the space densities of the main CV subclasses. Enlarged samples allow us to quantify the impact of magnetic fields on the evolution of CVs and also to synthesize the Galactic Ridge X-ray emission \citep[GRXE,][]{worrall+82}.

\begin{table*}[ht]
\centering
\caption{Cataclysmic variables detected as X-ray sources in eRO-DR1 and eRO-S45. The first column identifies the catalog or the type of CV as described in the text. The following columns give the grand total number of CVs in those catalogs, the number of those with \textit{\gai} DR3 information (column 3), the number of those with distances derived by \cite{bailer-jones+21} (column 4), the number of those that happen to lie in the western Galactic hemisphere (column 5), and the number of those within a radius of 500\,pc. The subsequent columns give the number and percentage of CVs that were detected as X-ray sources in eRASS1 (DR1) and in the stack of all eRASS' (S45), and within a radius of 500\,pc. Percentages were rounded to the nearest integer. 
}
\label{t:matches}
\begin{tabular}{lrrrrr|rr|rr|rr|rr|}
\hline\hline
Catalog & Tot & GDR3 & BJ21 & DE & R500 & \multicolumn{2}{|c|}{DR1} &  \multicolumn{2}{c|}{S45} & \multicolumn{2}{c|}{DR1, R500} & \multicolumn{2}{c|}{S45, R500}\\
& \# & \# & \# & \# & \# & \# & (\%) & \# & (\%) & \# & (\%) & \# & (\%) \\
(1) & (2) & (3) & (4)&(5)&(6)&(7)&(8)&(9)&(10)&(11)&(12)&(13)&(14)\\
\hline 
&&&&&&&&&&&&&\\[-1.5ex]
CBcat &      1429 & 1287& 1163&492 & 143 & 290 &  59 & 382 &  78 & 112 & 78 & 133 &  93 \\[1ex] 
DN-nomag &    735 & 633 & 560 &210 &  68 & 131 &  62 & 178 &  85 &  54 &  79 & 64 &  94 \\ 
DN-SU &       627 & 526 & 462 &168 &  54 &  95 &  57 & 139 &  83 &  40 &  76 & 50 &  93 \\
DN-UG &        85 &  84 &  75 & 36 &  11 &  30 &  83 &  33 &  92 &  11 & 100 & 11 & 100 \\ 
DN-ZC &        23 &  23 &  23 &  6 &   3 &   6 & 100 &   6 & 100 &   3 & 100 &  3 & 100 \\[1ex] 
NL-nomag &     93 &  91 &  87 & 35 &   9 &  23 &  66 &  29 &  83 &   8 &  89 &  8 &  89 \\ 
NL-SW &        20 &  20 &  20 & 10 &   3 &   5 &  50 &   9 &  90 &   3 &  60 &  3 & 100 \\ 
NL-UX &        44 &  43 &  39 & 17 &   4 &  11 &  65 &  12 &  71 &   3 &  75 &  3 &  75 \\ 
NL-VY &        29 &  28 &  28 &  8 &   2 &   7 &  88 &   8 & 100 &   2 & 100 &  2 & 100 \\ 
\hline
&&&&&&&&&&&&&\\[-1.5ex]
Polars &      218 & 208 & 188 &  92 &  47 &  59 &  64 &  80 &  87 &  37 &  79 &  45 &  96 \\[1ex] 
IPs (4--5 *) & 71 &  70 &  70 &  24 &   3 &  23 &  96 &  24 & 100 &   3 & 100 &   3 & 100 \\ 
IPs (2--3 *) & 79 &  66 &  61 &  32 &   3 &  23 &  72 &  29 &  91 &   1 &  33 &   3 & 100 \\ 
\hline
&&&&&&&&&&&&&\\[-1.5ex]
AM CVn     &    80 &  72 &  65 &  29 &  10 &  19 &  66 &  26 &  90 &   9 &  90 &  10 & 100 \\ 
Symbiotics &   315 & 306 & 292 & 144 &   1 &  15 &  10 &  20 &  13 &   1 & 100 &   1 & 100 \\[1ex] 
\hline
&&&&&&&&&&&&&\\[-1.5ex]
Pala CVs  &    42 &  42 & ---$^{1)}$ &  22 &  22 &  20 &  91 &  22 & 100 & --$^{2)}$ &  -- &  -- & -- \\ 
CRTS CVs  &   855 & 611 & 496 & 247 &  33 & 137 &  55 & 204 &  83 &  24 &  73 &  29 &  88 \\ 
SDSS CVs  &   591 & 527 & 469 & 148 &  34 &  69 &  47 & 112 &  76 &  22 &  65 &  28 &  82 \\[0.5ex] 
\hline
\end{tabular}\\
1) Given the short distance of the Pala CVs, we use inverted parallaxes; 2) Pala CVs are within 500 pc by definition
\end{table*}

We present the combined X-ray and optical properties of known CVs in the X-ray all-sky surveys performed with \ero. The samples used are assumed to cover the important subclasses of CVs, the main distinctions being made between nonmagnetic and magnetic CVs on the one hand side, and between long-period, high-mass-transfer systems and short-period, low-mass-transfer systems on the other side. We characterize the CV content of eRASS1, but we also describe the advantages gained through the stacking of all available \ero X-ray survey data. 

We are primarily interested in characterizing the standard CVs, that is, those with Roche-lobe-filling main sequence donors; but we also included closely related objects 
in the analysis, such as the AM Canum Venaticorum stars  (AM CVns or double degenerates (DDs)), and the Symbiotic Stars (SySts), which are~white dwarfs (WDs) or neutron stars that accrete matter from M giants via a stellar wind. We then further extended the analysis by characterizing mixed samples, which were composed using a special observing technique (photometrically from the Catalina Real-Time Transient Survey (CRTS) and from the Sloan Digital Sky Survey (SDSS)) or were drawn from a volume-limited sample within 150 pc \citep[the Pala sample;][]{pala+20}.

The paper is organized as follows. We first briefly describe the samples  used. We then present a collection of the relevant \textit{\gai} information and some additional X-ray information from \ero. We then describe the sample properties of the standard CVs, the related CVs, and the mixed samples based on the collected X-ray
and optical properties. This is done with the help of diagnostic diagrams and median sample parameters. Two subsections deal with the spectral properties of magnetic CVs and the presence and strength of soft blackbody-like components. We then present a discussion of our main findings and summarize our conclusions.

Throughout this paper we use \textit{\gai} coordinates from DR3 and, if available, distances ({\it rgeo)} derived by \cite{bailer-jones+21}. The used X-ray fluxes were taken from \ero catalogs: for DR1 the published values are used, and for S45 the catalog is based on processing version c020. These fluxes are given in the $0.2-2.3$\,keV  band, which is the band used for source detection. Fluxes were derived from the observed count rates by applying the standard energy conversion factor (ECF), which assumes an absorbed power-law spectrum. The usage of a more appropriate thermal plasma model would have resulted in lower fluxes by $5-10$\%.

\section{Sample analysis\label{s:obs}}
\subsection{\ero catalogs}
\ero has scanned the entire sky several times. The individual complete surveys are called eRASS1, eRASS2, eRASS3, and eRASS4. A fifth survey was completed to about 40\%. The results of eRASS1 were published recently in \citep{merloni+24} and we refer to it as DR1 (data release 1). DR1 comprises data, software, and a source catalog with more than 900,000 entries. The German \ero collaboration (\ede) has data rights for half of the celestial sphere, the western Galactic hemisphere. The exact definition is described in \cite{merloni+24}.
\ede has also generated a catalog based on the stacked data from all surveys which is not yet finally calibrated and accessible to the collaboration only. We make use of this resource here and refer to it in the following as S45.

\begin{figure*}
\resizebox{0.49\hsize}{!}{\includegraphics{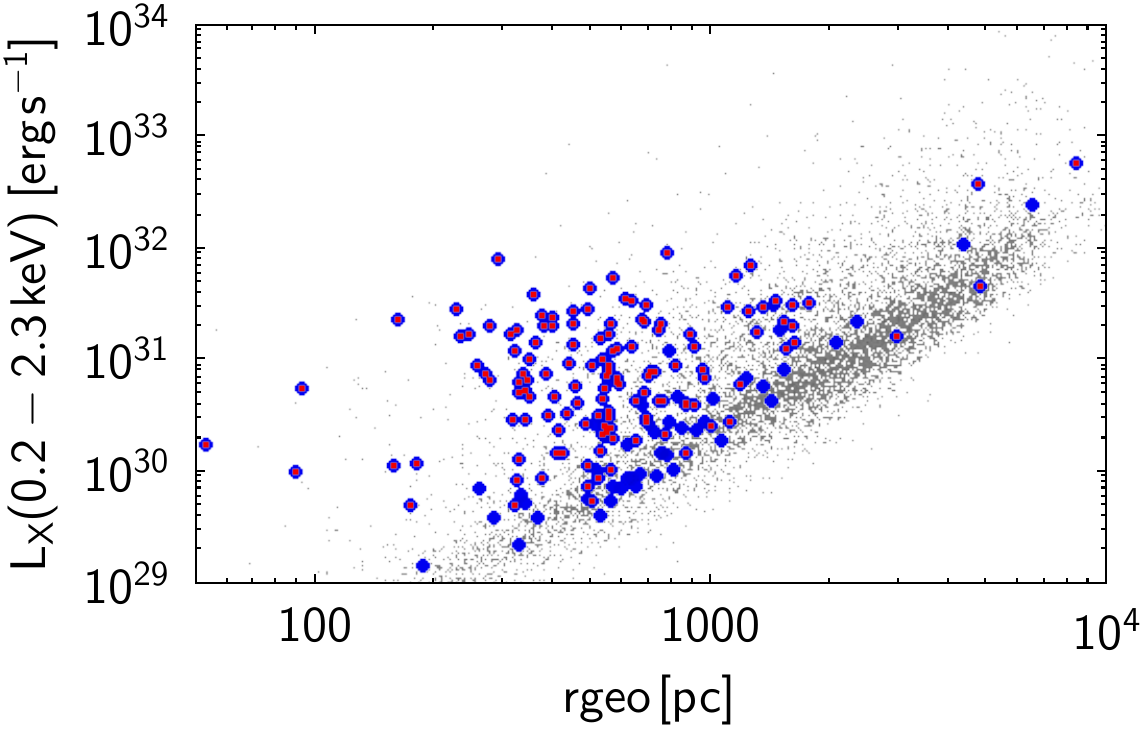}}
\hfill
\resizebox{0.49\hsize}{!}{\includegraphics{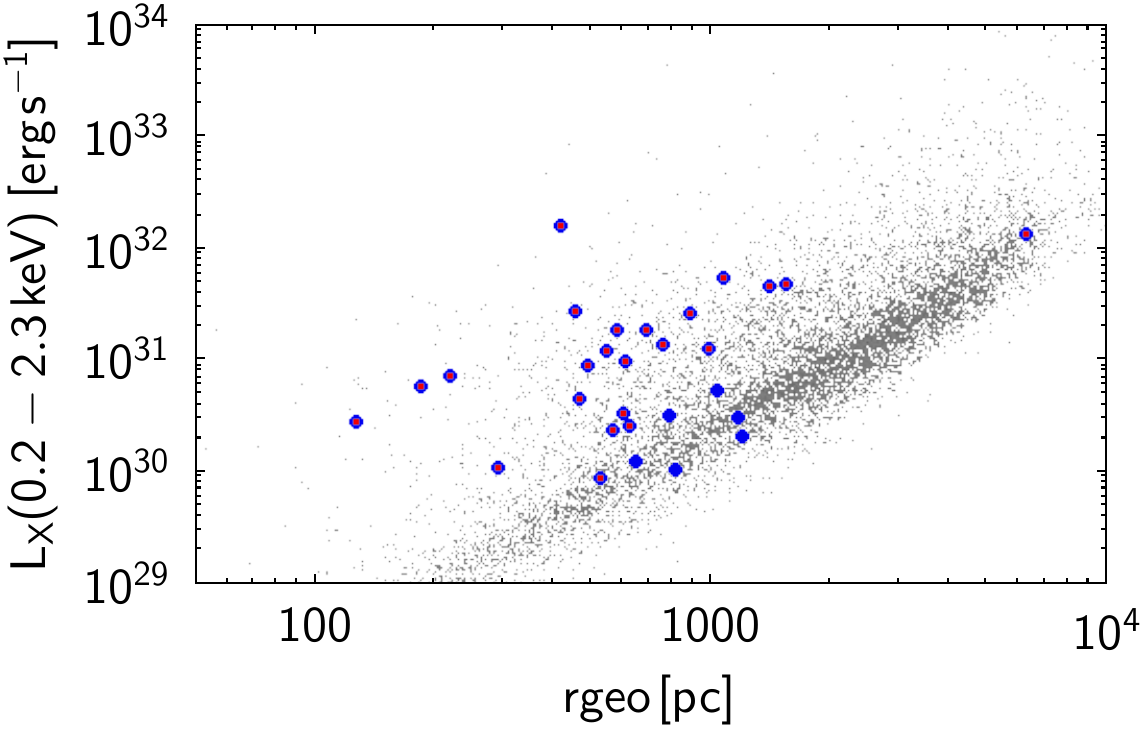}}
\resizebox{0.49\hsize}{!}{\includegraphics{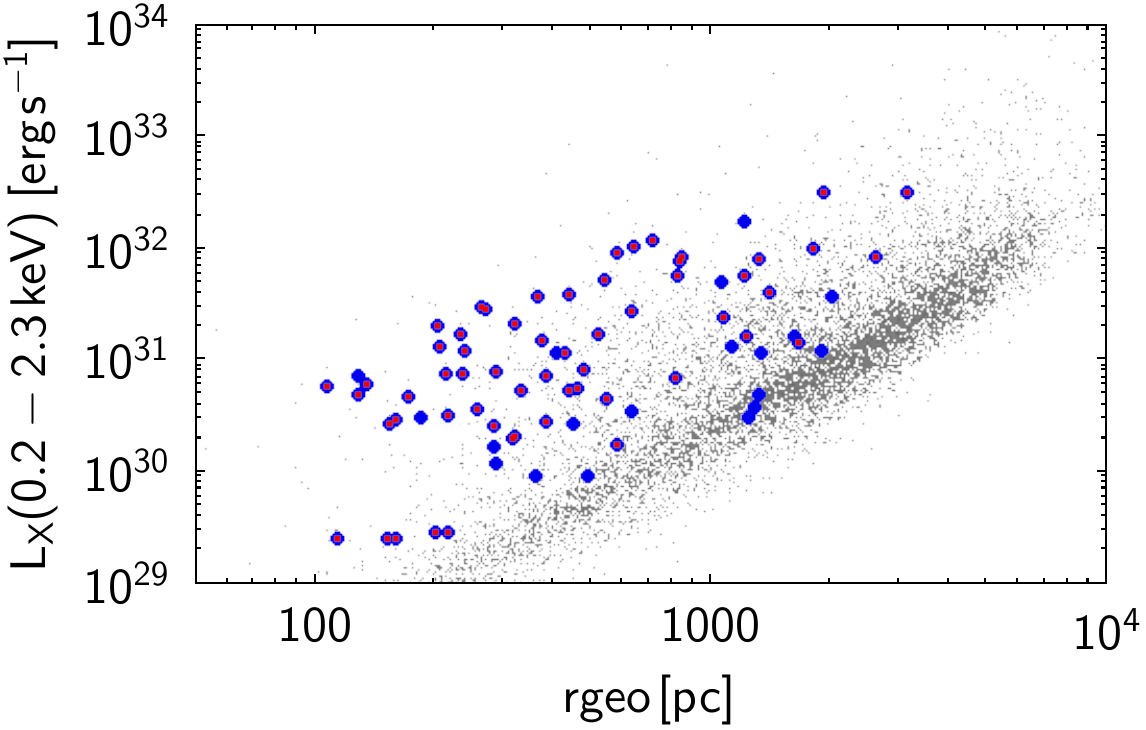}}
\hfill
\resizebox{0.49\hsize}{!}{\includegraphics{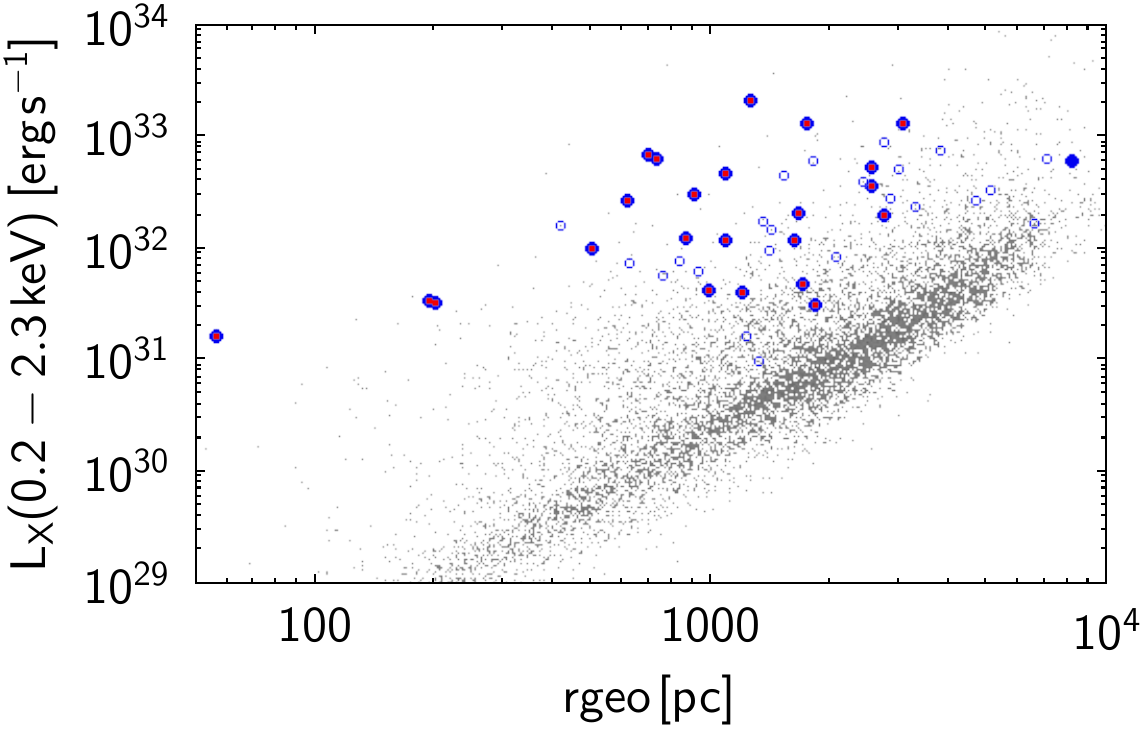}}
\caption{Location of the four main source classes of standard CVs detected in DR1 and S45 in the X-ray luminosity$-$distance plane. Blue filled symbols denote CV detections in S45, and those filled with a red dot denote objects also detected in DR1. Background objects shown with small gray symbols are made up of compact binary candidates based on eRASS:3 and objects chosen for spectroscopic follow-up in SDSS-V. The arrangement of panels here is consistent with the following four figures: nonmagnetic CVs are shown in the top row (DNe left, NLs right), magnetic CVs in the bottom row (polars left, IPs right).} 
\label{f:lxdist}
\end{figure*}

\subsection{Cataclysmic variable samples}
In the following paragraphs, we list the CV samples that were used to search for the X-ray detection of their members in DR1 and S45. For nonmagnetic CVs we use the last edition of the catalog of Cataclysmic Binaries, LMXBs, and Related Objects compiled by \cite{ritter_kolb03}\footnote{\url{http://cdsweb.u-strasbg.fr/cgi-bin/qcat?J/A+A/404/301}},
and refer to it as CBcat in the following. We use this catalog for its high reputation and wide use in the community despite the fact that \cite{inight+23a} have revised some of the classifications given therein. Of the 1429 entries in this catalog of all types of CVs, we use only the nonmagnetic CVs listed therein, the dwarf novae (DN; CBcat Type1==DN) with subtypes  SU UMa, U Gem, and Z Cam (CBcat Type2, abbreviated as SU, UG, ZC), and the novalikes (NL; CBcat Type1==NL) with subtypes SW Sex, UX UMa, and VY Scl (CBcat Type2, abbreviated as SW, UX, VY, respectively). We use only combinations of Type1 and Type2 to classify an object and ignore the Types 3 and 4 also given there. If an object was classified as Type1==DN, Type2==SU, and Type3==IP (like FS Aur), it would fall here in the CBcat DN category. 

The magnetic CVs are divided into two subgroups, the polars or AM Herculis stars and the intermediate polars (IPs). While the rotation of the WDs in the short-period polars is synchronized with the orbit, the long-period IPs have freely spinning WDs. The polars have no accretion disk, the IPs might have a truncated disk. In both groups, accretion happens quasi-radially via magnetically confined streams or curtains onto the polar regions of the WDs. The CBcat also lists magnetic CVs but the numbers are small, and so we use different resources. For IPs, we use Koji Mukai's list\footnote{\url{https://asd.gsfc.nasa.gov/Koji.Mukai/iphome/iphome.html}}, which was last updated in 2021. Mukai has assigned levels of reliability of an IP classification ranging from five-star ("ironclad") to one-star ({"doubtful"}) subgroups. We define two broader groups to not deal with excessively underpopulated subgroups and put those with four and five stars in one group and those with two and three stars in another. We ignore IP candidates with only one star. From the initial list of 202 objects, the most reliable subsample with four and five stars contains 71 objects, the less reliable subsample with two and three stars contains 79 objects. The sample of polars that is used here is our own collection, which has grown over the years to 218 objects and is based initially on CBcat entries with the addition of many individual discoveries reported in the literature. It was updated most recently at the end of 2023.

\begin{table*}[t]
\centering
\caption{Median parameters of CV samples derived for \textit{\gai} DR3 and S45. The first part of the table describes the properties of CV subclasses and the second part describes literature samples with a mixture of subclasses.}
\label{t:parms}
\begin{tabular}{lrccrrrc}
\hline\hline
    Class & $r_{\rm geo}$ & $\log L_{\rm X}$ & $\log(f_{\rm X}/f_{\rm opt})$ & HR1 & HR2 & $G_{\rm abs}$ &$G_{\rm BP}-G_{\rm RP}$\\
    & (pc) & (\lx) & & & & (mag) & (mag) \\
\hline
&&&&&&&\\[-1.5ex]
CBcat DN     &  565 & 30.75 & $-0.69$ & 0.67 &   0.09  & 9.6 & 0.69 \\ 
CBcat NL     &  624 & 30.85 & $-2.24$ & 0.69 & $-0.05$ & 5.4 & 0.30 \\[0.5ex] 
Polars       &  439 & 30.90 & $-0.32$ & 0.23 &   0.00  & 9.8 & 0.83 \\ 
IPs (4--5 *) & 1148 & 32.30 & $-0.61$ & 0.58 &   0.18  & 5.6 & 0.56 \\ 
IPs (2--3 *) & 1619 & 32.16 & $-0.81$ & 0.61 &   0.25  & 5.7 & 0.92 \\ 
&&&&&&&\\[-1.5ex]
AM CVns      &  737 & 30.61 & $-0.65$ & 0.52 &   0.06  &10.0 & 0.08\\ 
Symbiotics   & 2143 & 31.36 & $-3.68$ & 0.87 &   0.23  & $-1.6$ & 2.54\\ 
\hline
&&&&&&&\\[-1.5ex]
Pala CVs     &  117 & 30.34 & $-0.95$ & 0.52 & $-0.02$ & 10.6 & 0.43 \\ 
CRTS CVs     &  974 & 30.93 & $-0.56$ & 0.67 &   0.08  & 9.3 & 0.69 \\ 
SDSS CVs     &  818 & 30.88 & $-0.56$ & 0.57 &   0.06  & 9.3 & 0.60 \\ 
\hline
\end{tabular}
\end{table*}

Related to the standard CVs are the AM CVns and the SySts, which are also investigated here. Our sample of AM CVns is based on the compilation of the 56 objects from \cite{ramsay+18}. To those, we added objects that we became aware of through literature studies and our own work \citep[e.g.][]{khalil+24, schwope+24} so that the list of objects has grown to 80. Admittedly, the DDs are a heterogeneous class comprizing objects with rather different evolutionary pathways and different accretion scenarios, from direct impact accretors to NL- or DN-like objects. We nevertheless put them in one category because not for every object all the parameters are known and for not being forced to deal with too small input samples. For the SySts we use the compilation by \cite{akras+19} of galactic and extragalactic confirmed and candidate systems. We include all candidates but remove all extragalactic objects so that our input sample of SySts comprized 315 objects. These may include white-dwarf and neutron-star accretors. We made no attempt to revise the classifications given by Akras et al.~by e.g. making use of further literature studies, but see the notes below Table~\ref{t:symbs}.

A further class of related objects are the old novae. The scientific questions regarding those are different than for the normal CVs (in novae one is primarily interested in the progenitor objects, and the moment at which accretion resumes after a nova eruption), and therefore old novae are not investigated here. A separate study on on these objects is underway whose results will be presented separately (Sala et al., in preparation). 

As mentioned already, there are mixed samples of CVs described in the literature that are of considerable interest for the community. We also investigate some of the properties of three such samples in this work, namely the volume-limited sample of 42 CVs studied by \cite{pala+20}, the photometrically selected 855 CV candidates from the CRTS \citep{drake+14}, and the approximately 600 CVs from the SDSS, which are all confirmed as CVs through SDSS spectroscopy and further classified by \cite{inight+23a, inight+23b}. These are referred to as Pala CVs, CRTS CVs, and SDSS CVs, respectively, and are treated in the same manner as the other samples.

\begin{figure*}[t]
\resizebox{0.49\hsize}{!}{\includegraphics{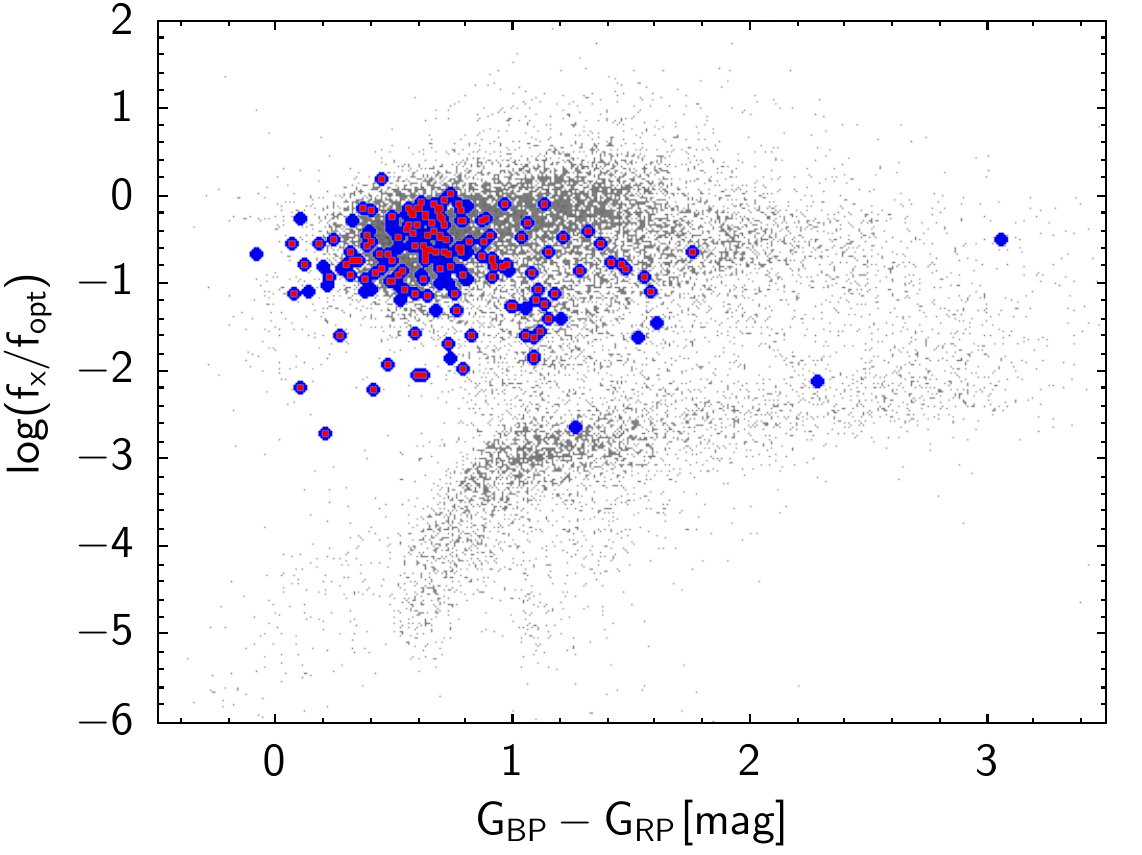}}
\hfill
\resizebox{0.49\hsize}{!}{\includegraphics{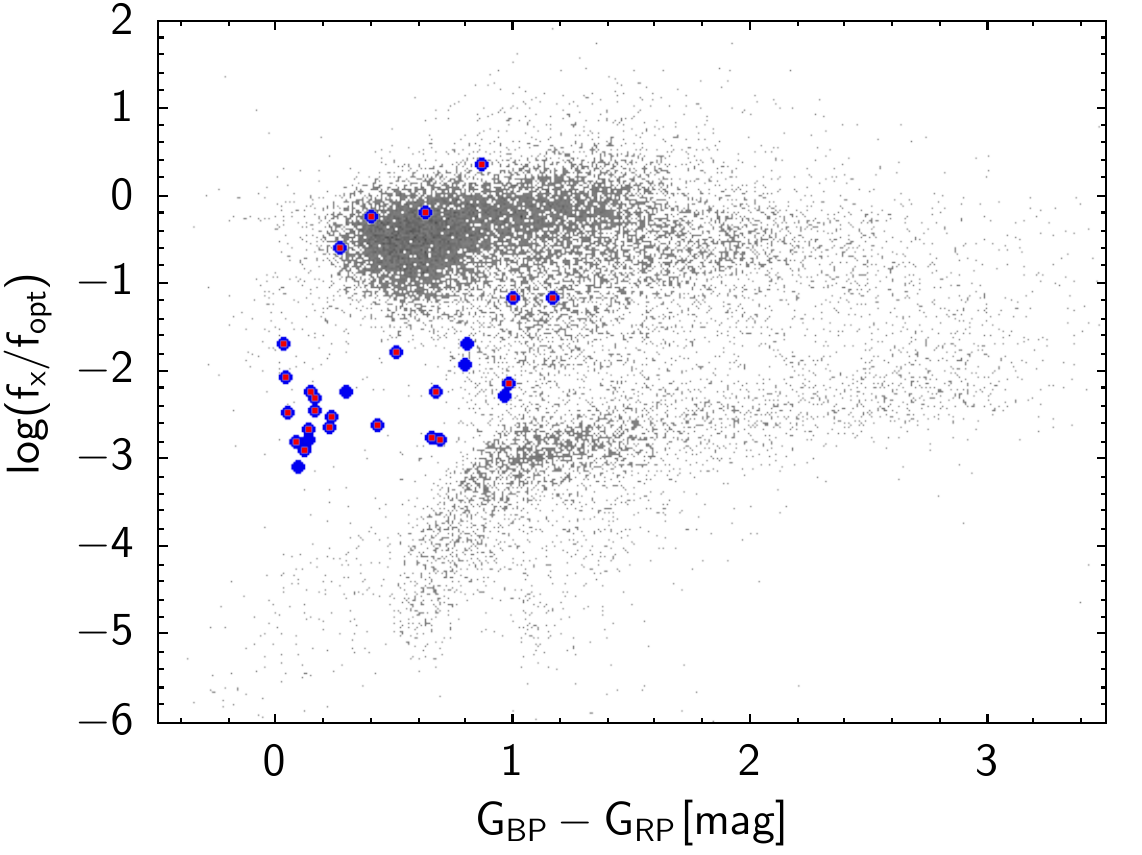}}
\resizebox{0.49\hsize}{!}{\includegraphics{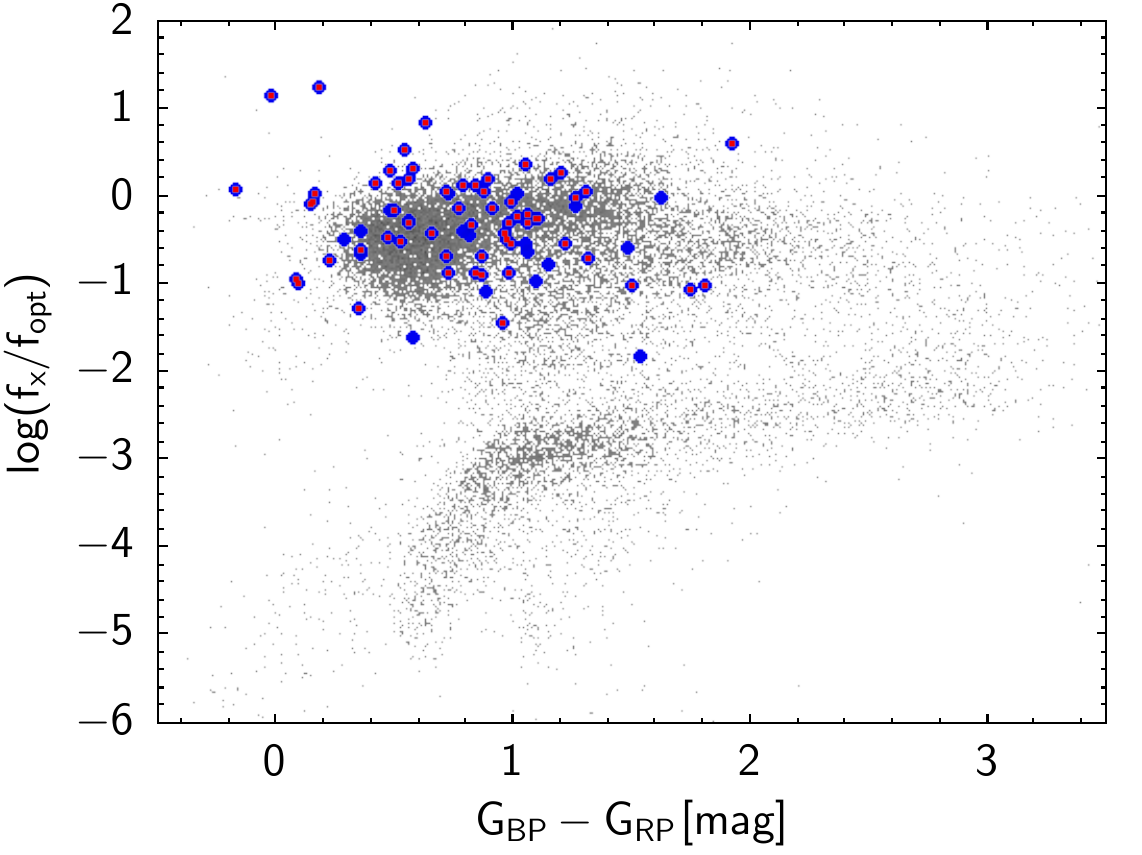}}
\hfill
\resizebox{0.49\hsize}{!}{\includegraphics{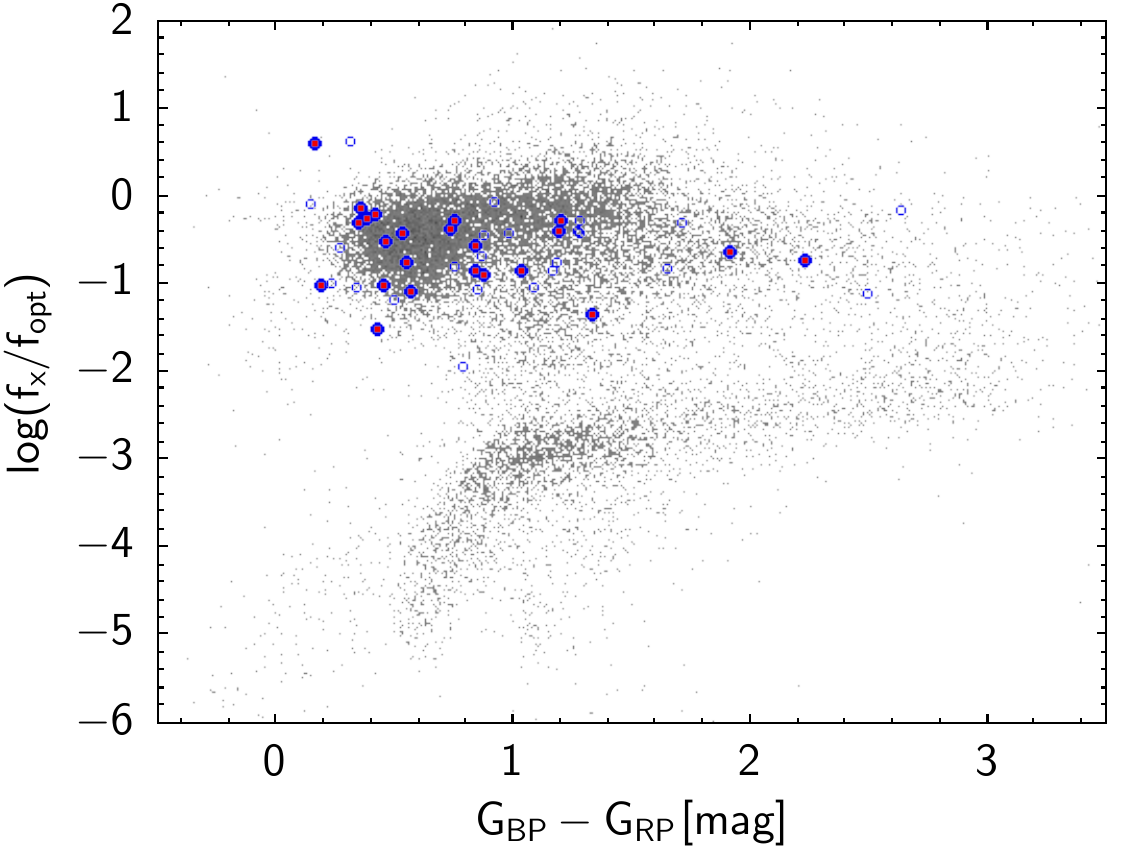}}
\caption{X-ray to optical color--color diagram of standard CVs detected in DR1 and S45. The color scheme is the same as in Fig.~1. The background is made of \textit{\gai} sources that match eRASS:3 sources within 1 arcsec.} 
\label{f:ccd}
\end{figure*}

\begin{figure*}[t]
\resizebox{0.49\hsize}{!}{\includegraphics{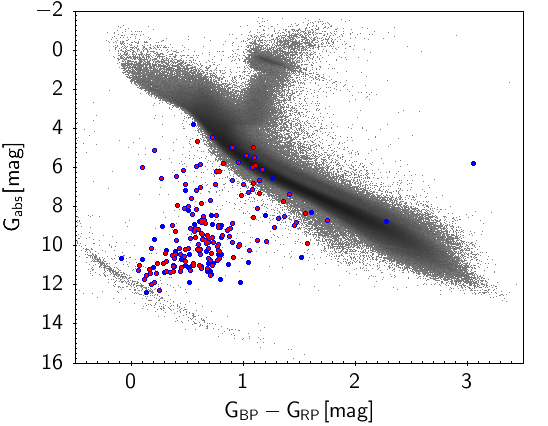}}
\hfill
\resizebox{0.49\hsize}{!}{\includegraphics{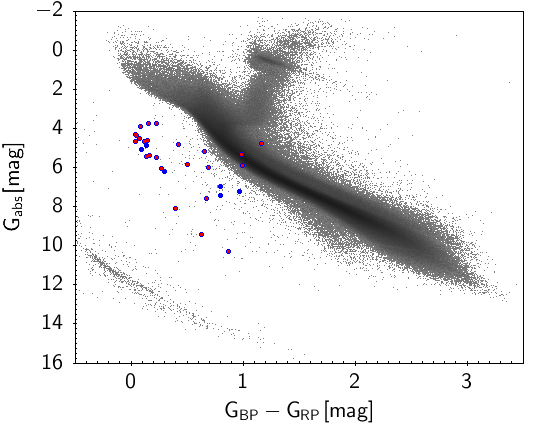}}
\resizebox{0.49\hsize}{!}{\includegraphics{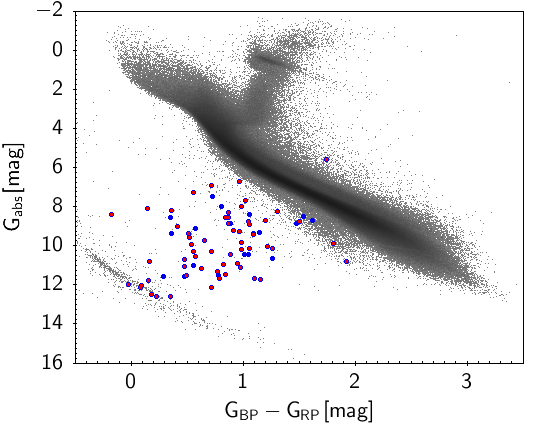}}
\hfill
\resizebox{0.49\hsize}{!}{\includegraphics{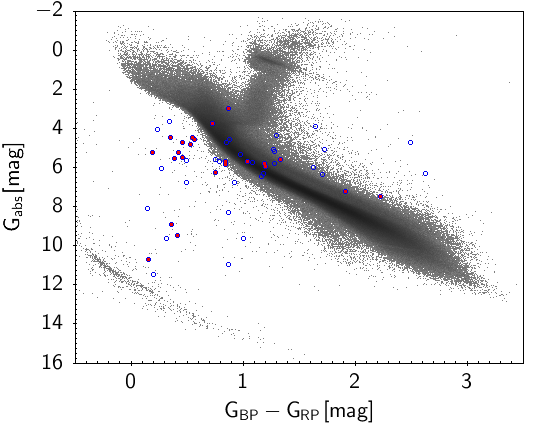}}
\caption{Color--magnitude diagram of standard CVs detected in DR1 and S45. The color scheme is the same as in Fig.~1. The background is made of nearby \textit{\gai} ($D_{\max} = 300$\,pc) sources with small parallax errors ($\pi > 33 \mbox{\,mas}, \Delta\pi<0.05$\,mas. Distances for these background objects were computed as inverted parallaxes.}
\label{f:cmd}
\end{figure*}

\begin{figure*}[t]
\resizebox{0.49\hsize}{!}{\includegraphics{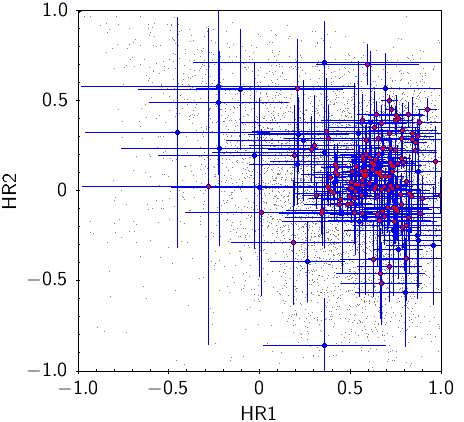}}
\hfill
\resizebox{0.49\hsize}{!}{\includegraphics{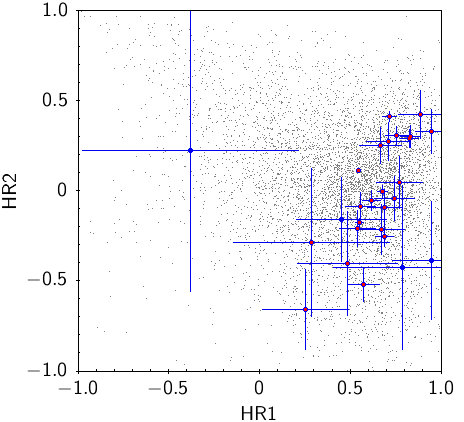}}
\resizebox{0.49\hsize}{!}{\includegraphics{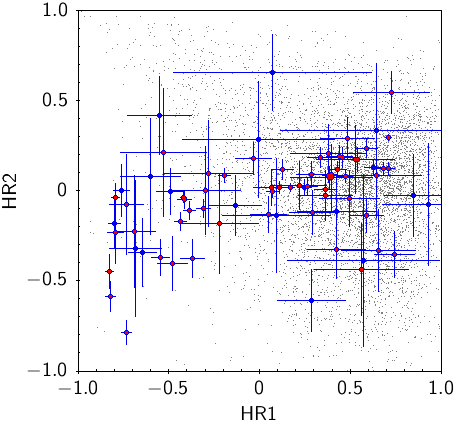}}
\hfill
\resizebox{0.49\hsize}{!}{\includegraphics{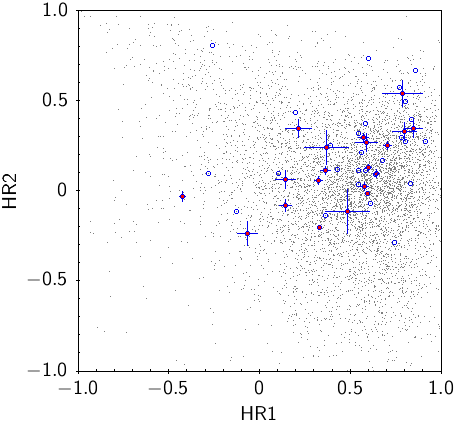}}
\caption{X-ray spectral hardness ratios of standard CVs detected in DR1 and S45. The color scheme is the same as in Fig.~1.}
\label{f:hrs}
\end{figure*}

\begin{figure*}[t]
\resizebox{0.49\hsize}{!}{\includegraphics{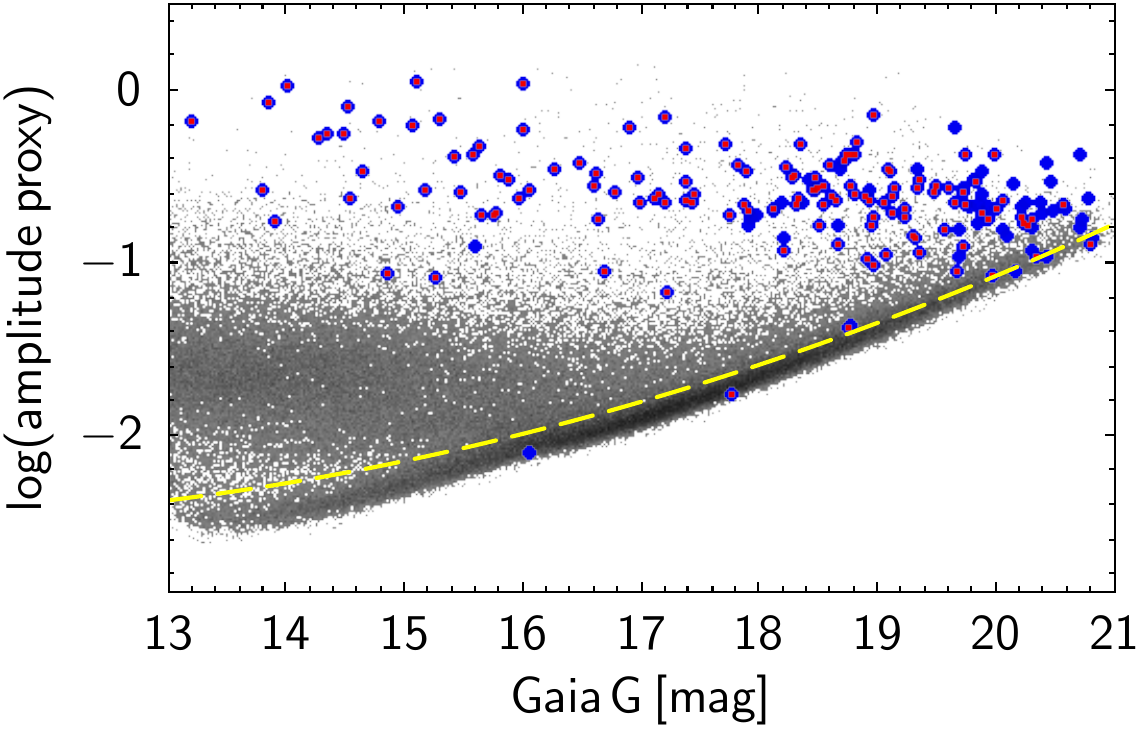}}
\hfill
\resizebox{0.49\hsize}{!}{\includegraphics{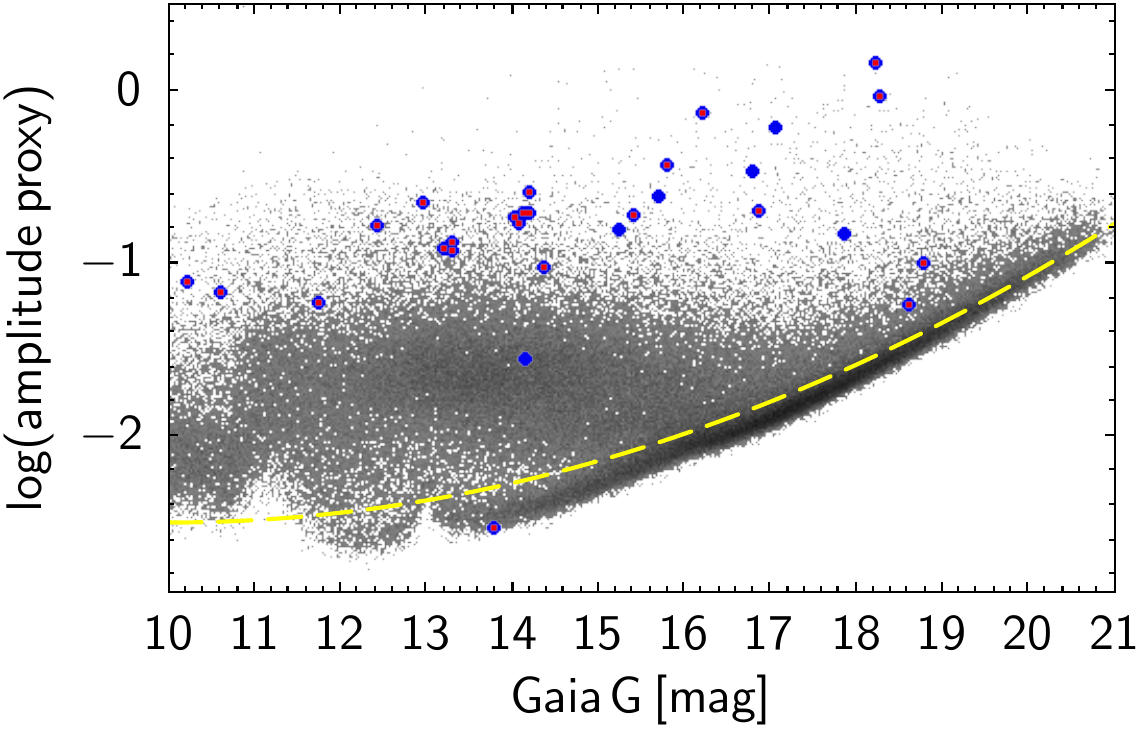}}
\resizebox{0.49\hsize}{!}{\includegraphics{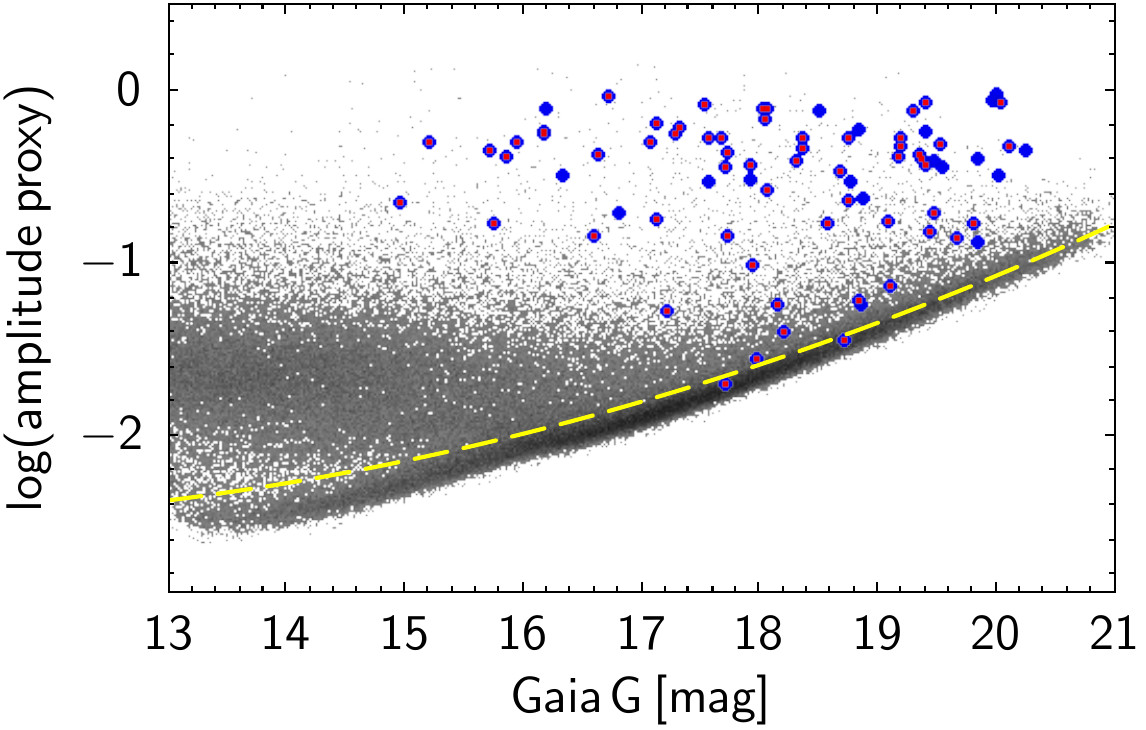}}
\hfill
\resizebox{0.49\hsize}{!}{\includegraphics{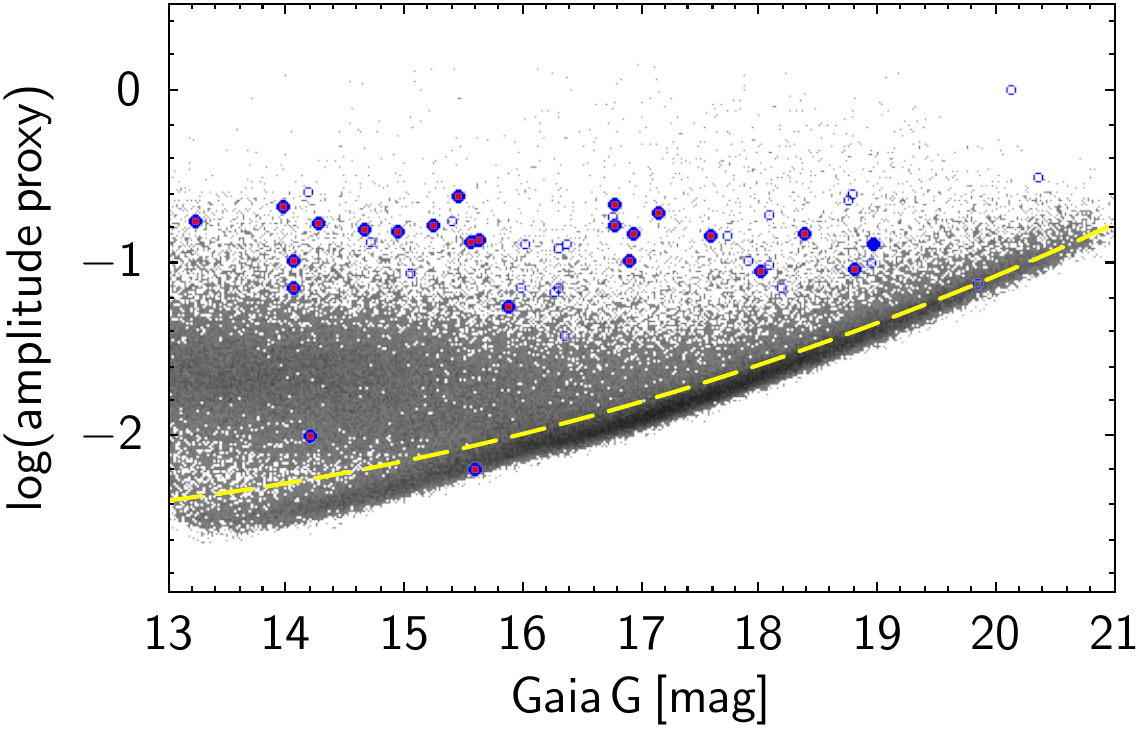}}
\caption{Variability parameter of standard CVs as determined by \cite{eyer+20}. The color scheme is the same as in Fig.~1. We note the different $x$-scale for the NLs.}
\label{f:var}
\end{figure*}

\subsection{Cataclysmic variables detected as \ero X-ray sources}
To investigate the combined X-ray and optical properties, we first needed \textit{\gai} coordinates for our systems. We used TOPCAT and STILTS \citep{taylor05} to crossmatch the initial lists of CVs with \textit{\gai} DR3 and then derive distances from \cite{bailer-jones+21}. The samples used by us were collected from various sources with unknown positional accuracy. For the initial crossmatch between the input lists and \textit{\gai} DR3, we accepted matches within 3 arcsec and considered matches at larger radii as potentially spurious. Of the initial 1429 CVs in the CBcat, we found 1287 with \textit{\gai} coordinates, of which 1163 have distances from \cite{bailer-jones+21}. Of those, 492 are located at galactic longitude $l^{II}>180\degr$ and 143 of those are closer than 500\,pc. All those numbers derived for the various CV samples are listed in Table~\ref{t:matches}. The first six columns identify the sample, give the total number (Tot), those with \textit{\gai} photometry (GDR3) and distances from \citep[BJ21][]{bailer-jones+21}, of those that fall in the 'German' hemisphere (DE), and of those that are closer than 500 pc (R500). The last column is generated to sketch the potential of a volume-limited search for CVs based on X-ray selection. For comparison, the volume-limited sample studied by \cite{pala+20} with a distance cut at 150\,pc has 42 CVs (of which 22 lie in the western galactic hemisphere). It thus suffers from a small sample size, one needs significantly larger samples to overcome the Poisson noise. The chosen distance limit of 500\,pc corresponds to an X-ray luminosity of $L_{\rm X} \mbox{(0.2-2.3 keV)} \sim 1.5 \times 10^{30}$\,\lx at the DR1 flux limit and of $\sim 6 \times 10^{29}$\,\lx at the S45 flux limit, respectively,. These are at about the right limit to even find period bouncing CVs \citep{munoz-giraldo+23, munoz-giraldo+24} and Low-Accretion Rate Polars  \citep[LARPs][]{schwope+02b}. Table~\ref{t:matches} has four sections. The top lists all entries from the CBcat and the nonmagnetic CVs listed therein, the second section the magnetic CVs, the third section the related systems (AM CVns and SySts), and the bottom section the mixed samples from Pala, the SDSS, and the CRTS. 

We then searched for entries in the two \ero catalogs DR1 and S45. Again we used TOPCAT/STILTS for the best positional match within 3$\sigma$ positional uncertainty of the X-ray positions for DR1 (\texttt{POS\_ERR} in the DR1 paper and the corresponding catalog). The relation between the statistical uncertainty (\texttt{RADEC\_ERR}) and the final positional uncertainty was calibrated for DR1 \citep{merloni+24} but is not yet available for S45. We therefore use for S45 the same formula but apply a more generous 5$\sigma$ limit. The \textit{\gai} coordinates have negligible positional uncertainty compared to the X-ray data. The number of matches thus found are documented in the 7th and the 9th columns of Table~\ref{t:matches}, respectively. Percentages with respect to column (5) are also given there. The last four columns of Table~\ref{t:matches} give the numbers and percentages of matching objects that lie within 500\,pc, again to sketch the opportunities and limitations of a volumetric census. 

The fraction of CVs for which an \ero counterpart was found lies typically around 65\% for DR1 (column 8) and between 80\% to 90\% for S45 (column 10). For DR1 there are three subclasses that stick out, the DN-ZC, the NL-VY, and the IPs with percentages 100\%, 88\%, and 100\%, respectively. However, all have rather few members and are within 1$\sigma$ of the average behavior of the other subsamples.

The first row in Table~\ref{t:matches} lists the results for all entries in the CBcat. Of the initial 1429 objects about one-third have a \textit{\gai} distance and lie in the Western Galactic hemisphere. Of those about 30\% are within 500\,pc. Most of them (92\%) can be recovered in S45, but the distribution over source classes is a bit sobering. There are just 64 DN found in S45 and 8 nonmagnetic NLs. Any statistical assessment of this latter sample is subject to Poisson noise. The polars were discovered in significant numbers in the RASS. This sample is the second most numerous within 500 pc after the DN-SU. A fraction of 64\% of the known sample was detected in DR1, the fraction rises to 96\% within 500\,pc in S45. Almost all the secure IPs in the western Galactic hemisphere with known distances are contained in DR1, but very few are found within 500\,pc. An IP survey needs a large volume. Fortunately, the IPs are luminous, which facilitates such an attempt (see below). More than 90\% of Pala's CVs were detected in DR1. The two objects that were not detected in DR1 are V379 Tel and \textit{Gaia} J154008.28$-$392917.6. The former has an eROSITA upper limit flux of $1.2 \times 10^{-13}$\,\fergs for a vignetting corrected exposure time of 85.7\,s \citep{tubin+24}. In S45 it has a mean flux of $3.55 \times 10^{-12}$\,\fergs and it is the fourth brightest object among the 22 that are in the western Galactic hemisphere, indicating that strong source variability has prevented its discovery in eRASS1. For \textit{Gaia} J154008.28$-$392917.6 the vignetting-corrected exposure was 112.6 s and the upper limit flux was $1.57\times 10^{-13}$\,\fergs in DR1. The object was detected in S45 at a level of $8.9\times 10^{-14}$\fergs.

Only about half of the CRTS and SDSS CVs that lie in the western Galactic hemisphere and have known distances are detected in DR1. This fraction is raised to about 80\% in S45. Both samples are rather distant (see the next section), and therefore they are rare in a volume with a radius of 500\,pc. 

The discussion of the DDs and the SySts is deferred to later subsections (Sects.~\ref{s:amcvns} and \ref{s:systs})

\subsection{Properties of the cataclysmic variable samples }
With combined X-ray and optical properties established, we may look into sample properties. We did so by determining average optical and X-ray parameters. We list those sample properties in Table~\ref{t:parms} and constructed diagnostic scatter diagrams for all class members. In Table~\ref{t:parms}, we list median values for the distance, the X-ray luminosity, the ratio between the X-ray and the optical flux, the X-ray hardness ratios HR1 and HR2, the absolute magnitude in \textit{\gai} $G$-band, and the optical color $G_{\rm BP} - G_{\rm RP}$. Proper definitions of those quantities are given in the following paragraphs. The given values were derived from the sample of CVs detected in S45. The diagnostic diagrams are the color--magnitude diagram (CMD), an X-ray to optical color--color diagram (CCD), an X-ray color--color (hardness ratio) diagram, a luminosity--distance diagram, and a variability diagram. The diagrams shown in Figs.~\ref{f:lxdist} to \ref{f:var} were each made for the four main subclasses, the nonmagnetic DN and NLs, and the magnetic polars and IPs. The sample properties and diagnostic diagrams for the nonstandard CVs, the AM CVns, and the SySts are discussed in separate subsections below.

The X-ray to optical flux ratio for the CCD (Fig.~\ref{f:ccd}) was computed as $\log{(f_{\rm X}/f_{\rm opt})} = \log{f_{\rm X}} + \mbox{phot\_g\_mean\_mag} / 2.5 + 4.86$. The zeropoint was found by folding a flat spectrum with mag$_{\rm AB} = 0$ through the \textit{Gaia} passband\footnote{\url{https://www.cosmos.esa.int/web/gaia/iow_20180316}} followed by an integration. 
The X-ray luminosity was calculated as $4 \pi r^2_{\rm geo}f_{\rm X}$ for all the CVs, although the geometry factor for the magnetic CVs is likely a factor 2 lower, but may be different also from system to system. All luminosity values are based on the observed X-ray fluxes, not corrected for interstellar absorption. The impact is small for nearby objects. The flux in the detection band (0.2-2.3 keV) of a 30 keV thermal spectrum is reduced to 94\% (69\%) of its unabsorbed level for a column of cold interstellar matter with $N_{\rm H} = 1 \times 10^{20}$\,cm$^{-2}$ ($1\times 10^{21}$\,cm$^{-2}$). The latter high absorption value was found as approximate maximum from spectral fits to the soft IPs (see Table\,\ref{t:ips}).  
The absolute magnitude shown in Fig.~\ref{f:cmd} was computed using \gai $G$ magnitudes and the distance $r_{\rm geo}$ from \cite{bailer-jones+21}. Again, extinction from \gai\  was not taken into account to compute absolute magnitudes, because of our insecurity on the reliability of the given values for CVs, whose intrinsic colors are not known and, furthermore, strongly variable with accretion state and orbital phase. The source class, where extinction is very important, are the SySts (see Sect.~\ref{s:systs}). X-ray spectral hardness (Fig.~\ref{f:hrs}) is defined as HR$=\frac{H-S}{H+S}$ where $H,S$ are counts in a hard and a soft band, respectively. HR1 involves the bands $S = (0.2-0.5)$\,keV, $H = (0.5-1.0)$\,keV, and HR2 the bands $S = (0.5-1.0)$\,keV, $H = (1.0-2.0)$\,keV. 

The amplitude proxy shown in Fig.~\ref{f:var} as a function of the apparent G-band brightness was determined following \citep[][ their Fig.~1]{eyer+20}. They used the uncertainty on the mean and the number of per-CCD measurements to estimate the standard deviation of the (unpublished) photometric time series. The logarithm of the amplitude proxy was thus calculated as $\log(\mbox{sqrt}(\mbox{phot\_g\_n\_obs)/phot\_g\_mean\_flux\_over\_error})$. 
The yellow line in the variability diagrams (Fig.~\ref{f:var}) was calibrated by \cite{eyer+20} and indicates an average for constant objects. 

The arrangement of panels in the figures for the standard CVs is always the same. The upper row shows the nonmagnetic CVs (DNe left, NLs right) and the bottom row shows the magnetic CVs (polars left, IPs right). Smaller red symbols are for DR1 and bigger filled blue symbols for S45. For the IPs, this schema was used for objects with 4 and 5 stars. IPs with 2 and 3 stars in S45 are shown with open blue circles in all the panels.

The various diagrams have a background of sources that shall guide the eye and put the CVs from the various subsamples in context with other CV candidates or other unrelated objects, like stars or AGN. For the luminosity-distance, the hardness ratio, and the variability diagrams (Figs.~\ref{f:lxdist}, \ref{f:hrs}, and \ref{f:var}) we use the candidate objects that were submitted for spectroscopic follow-up to the SDSS-V collaboration as background objects. For the CMD (Fig.~\ref{f:cmd}) we used a random sample of \textit{\gai} objects within 300 pc, for the CCD (Fig.~\ref{f:ccd}) we used \textit{\gai} objects that were positionally matched to \ero sources drawn from eRASS:3 within 1 arcsec. The \textit{\gai} objects in the CMD trace the white-dwarf and the main sequences. The matched \textit{\gai}/\ero objects used in the CCD fall into two main groups, the AGN at reasonably blue color and high \fxo, and the stellar coronal emitters stretching from (log(\fxo),B-R)=($-$5.5,0.6) to ($-$2.5,3.2). The CMDs shown here may be compared to those shown by \cite{abril+20}, who constructed CMDs for all the CBcat CVs and their subclasses, not paying attention to X-ray selection.

The various subclasses of CVs analyzed here show a significantly differing behavior of their directly observed (optical color, \fxo, X-ray hardness, variability) or derived properties (distance, $L_{\rm X}$, $G_{\rm abs}$). The large dispersion is explained by a mixture of reasons, intrinsic properties, and detection biases playing important roles. The first two columns of \tpar list the median distance $r_{\rm geo}$ and the X-ray luminosity $L_{\rm X}$ for all subsamples. Scatter plots for these quantities are shown in Fig.~\ref{f:lxdist}. The other columns in \tpar list median values of the X-ray to optical flux ratio, the hardness ratios HR1 and HR2, the absolute magnitude $G_{\rm abs}$ and the optical color $G_{\rm BP} - G_{\rm RP}$. 

\subsubsection{Luminosity--distance diagrams} 
The IPs represent, by a large factor, the most luminous subclass. This is expected and well-known but was not yet demonstrated for \ero. Almost none of the IPs is found at the detection limit, all known systems are found by several orders of magnitude above the limit (which is delineated by the bulk of background objects in Fig.~\ref{f:lxdist}), indicating that a significant population of low-luminosity IPs and/or distant IPs is yet to be discovered. Confirmation of those will be time-consuming because one needs to find the spin- and orbital periods independently in time-resolved X-ray data. The slightly less luminous IP candidates (those with assigned 2 and 3 stars, see Table \ref{t:parms}) are even more distant at $\langle r_{\rm geo}\rangle \sim 1600$\,pc and are thus even more difficult to classify. The paucity of high-luminosity IPs within 500\,pc may be considered as direct proof of their low space density. Only a few CV candidates reach a high X-ray luminosity of the order of $\sim$10$^{33}$\,\lx, as found for the few bright classical IPs.

Their magnetic cousins, the polars, represent the most local sample although their X-ray luminosity is of the same order as that of the DNe and the NLs. Their X-ray luminosity varies by a factor of 100 at a given distance (approximately the same factor as for the IPs) indicating a large scatter in mass accretion rates plus the presence of a variety of different energy release channels. High-B systems tend to have reduced X-ray fluxes from the cooling plasma \citep[see e.g.][and references given there]{traulsen+15}. 
They seem to show a flat distribution between 0.1 and 3\,kpc (contrary e.g. to the DNe, which shows a more concentrated distribution between 0.2 and 2\,kpc). Polars are detected down to the detection limit of \ero. Their preponderance at short distances is likely due to their most extreme \fxo values, and their soft X-ray spectra (the softest of all CV subclasses probed here), which made them prime candidates for optical follow-up of soft X-ray missions like \ros and \eins. 

\subsubsection{Color--magnitude diagrams} 
The DNe and the polars are the two subclasses with low optical brightness that fill the space between the white-dwarf and the main sequence in the CMD (cf.~Fig.~\ref{f:cmd}) with some overlap with both sequencies. NLs and IPs have large absolute magnitudes and a considerable overlap with the main sequence in their CMDs, the nonmagnetic NLs less than the IPS due to their extreme blue color at similar absolute magnitude.

\subsubsection{X-ray and optical color--color diagrams}
The flux ratio \fxo is highest for the polars and lowest for the NLs (Fig.~\ref{f:ccd} and column in Table~\ref{t:parms} ). Their optical colors are also the most extreme. NLs are extremely blue, while polars are very red. The high ratio \fxo of the polars is due to pure radial accretion without an accretion disk, while the bright, hot disks in NLs cause the blue optical color and reduce the X-ray to optical flux ratio. The red optical color of the polars is due to cyclotron radiation and photospheric radiation from the late-type donor, which is often seen in the spectra of polars. The \fxo for the DNe and IPs is similar and also somewhat below the main AGN blob in the CCD, which is centered on $\sim -0.5$. Their slightly smaller \fxo compared to the polars is due to the presence of accretion disks in these systems, which redirect some of the gravitational energy into the optical channel. There is almost zero overlap 
of any subsample with the stellar locus in the CCDs.

\subsubsection{X-ray hardness ratios}
The X-ray spectra can be characterized by the X-ray hardness ratio diagrams (Fig.~\ref{f:hrs} and columns 5 \& 6 in \tpar). XSPEC \citep{arnaud96} models show that a mildly absorbed thermal plasma with $kT>3$\,keV has HR1$\simeq 0.35$ and HR2$\simeq 0.1$. The chosen energy bands to determine hardness ratios are insensitive to higher temperatures. They were chosen to uncover possible multi-component spectra; of particular interest are additional soft components in the magnetic CVs. The hardness ratios HR1 and HR2 of all the DNe are compatible with thermal plasma emission with temperatures higher than $\sim$3\,keV. The NLs show a large spread in HR2, the observed distribution of the HRs is still compatible with thermal plasma emission. The negative HR2 indicates low plasma temperatures of order $1-2$\,keV. 

The HRs of the magnetic CVs are different. HR2 is mostly compatible with the thermal spectra from the cooling accretion plasma with temperatures in excess of 5 keV, while HR1 indicates the presence of a soft component, which is particularly prominent for the polars. If a cut is made at HR1$+\Delta$HR1$<0.35$ to define systems with a soft component, 37 out of 78 polars in S45 are included, i.e. almost 50\%. Most of these sources were discovered with \ros \citep[e.g.][]{thomas+98}.
The high fraction of soft polars is thus related to a detection bias. Most if not all of the \xmmn-discovered and also the first SRG-discovered polars lack the soft component \citep[see e.g.][and references therein]{ok+23}. One may reasonably expect that this detection bias will be overcome with the \ero follow-up, because the detection band between 0.2 and 2.3 keV is dominated by the thermal plasma emission from the accretion column and not by the highly instationary blobby accretion feeding the soft component. In Sect.~\ref{s:polars} we quantify the soft and hard X-ray components for bright polars, i.e. those with more than 500 photons.

HR2 of the IPs is compatible with hot thermal plasma emission, as expected for these objects. Applying the same criterion for selecting soft emitters to the secure IPs (Mukai's IPs with 4 and 5 stars) as for the polars, HR1$+\Delta$HR1$<0.35$, 7 out of 24 are selected (29\%). This fraction is significant. \cite{anzolin+08} report the discovery of soft components in \xmmn spectra of two \ros-discovered IPs and present a list of 13 soft IPs, i.e. those with a soft component in their spectra. Their fraction of "soft IPs" is 42\%, even higher than reported here. Of their 13 objects, only seven are in the western Galactic hemisphere. Of these, the value of HR1 of V2400 Oph, EX Hya, and WX Pyx does not indicate a soft component. Hence, four systems are common between our and their list of soft IPs (UU Col, PQ Gem, V418 Gem, NY Lup) while our HR1 criterion indicates another three objects not listed by them (TX Col, BG CMi, V667 Pup). All seven objects selected here have sufficient counts to allow for a spectral analysis. Details are reported in  Sect.~\ref{s:softips} but can be summarized as follows: All selected IPs require a soft component for a successful fit. The soft component in TX Col, BG CMi, and V667 Pup is reported here for the first time. The temperature of the assumed blackbody varied between 47\,eV for PQ Gem at the low end and 96\,eV at the high end for NY Lup and TX Col. The high temperature of NY Lup is compatible with earlier reports by \cite{haberl+02} (deriving $88\pm2$\,eV) and \cite{evans_hellier07} (deriving $104^{+21}_{-23}$\,eV) from \xmmn. Also, the low temperature of PQ Gem is the same as found with \xmmn by \cite{evans_hellier07}, who derive $47.6^{+2.9}_{-1.4}$\,eV.

\begin{figure}
\resizebox{\hsize}{!}{\includegraphics{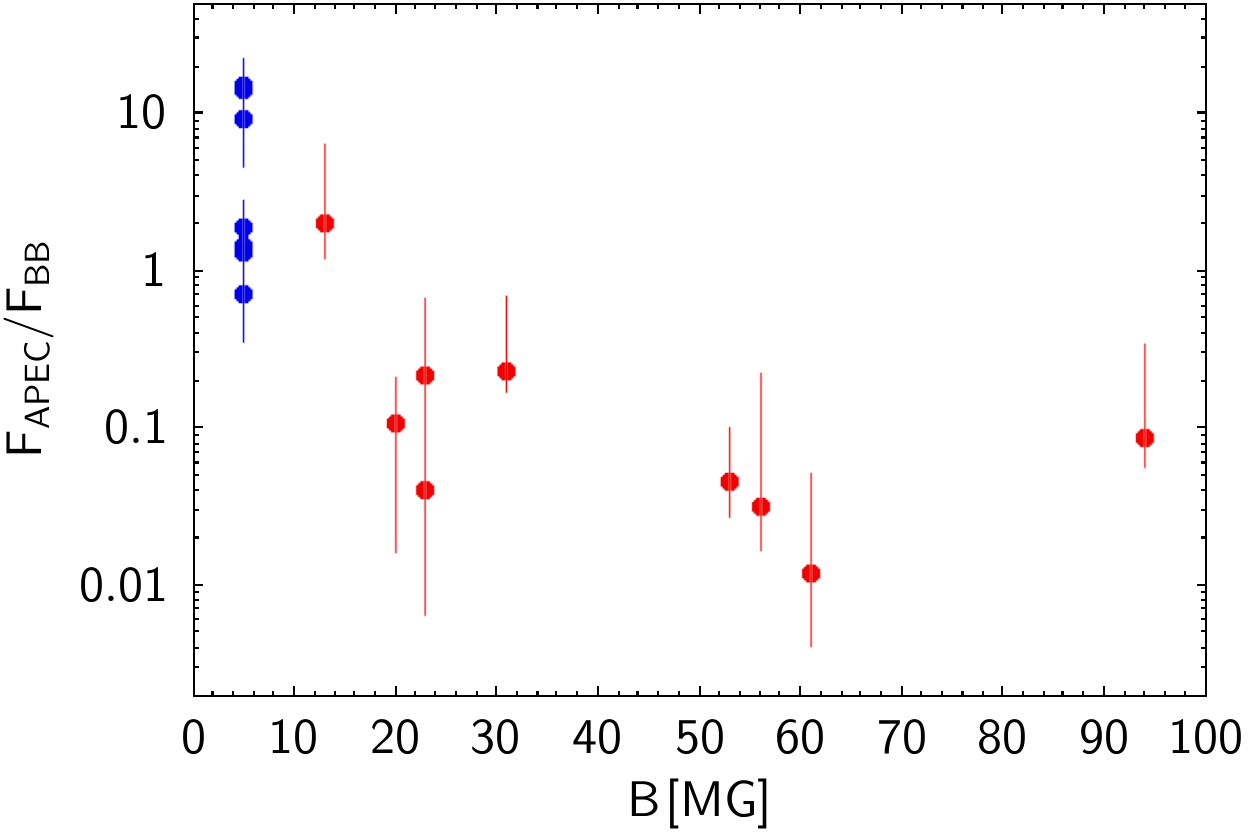}}
\caption{Ratio of the bolometric energy fluxes $F_{\rm APEC}/F_{\rm BB}$ as a function of the field strength $B$ for polars (red symbols) and for secure IPs (blue symbols). The temperature of the optically thin thermal spectral component was fixed at 15 keV for both classes, and the blackbody temperatures and fluxes are listed in Tables~\ref{t:polars} and \ref{t:ips}. Flux errors were determined from the 90\% confidence regions of the normalization constants.}
\label{f:polsoft}
\end{figure}

\begin{figure*}
\resizebox{0.49\hsize}{!}{\includegraphics{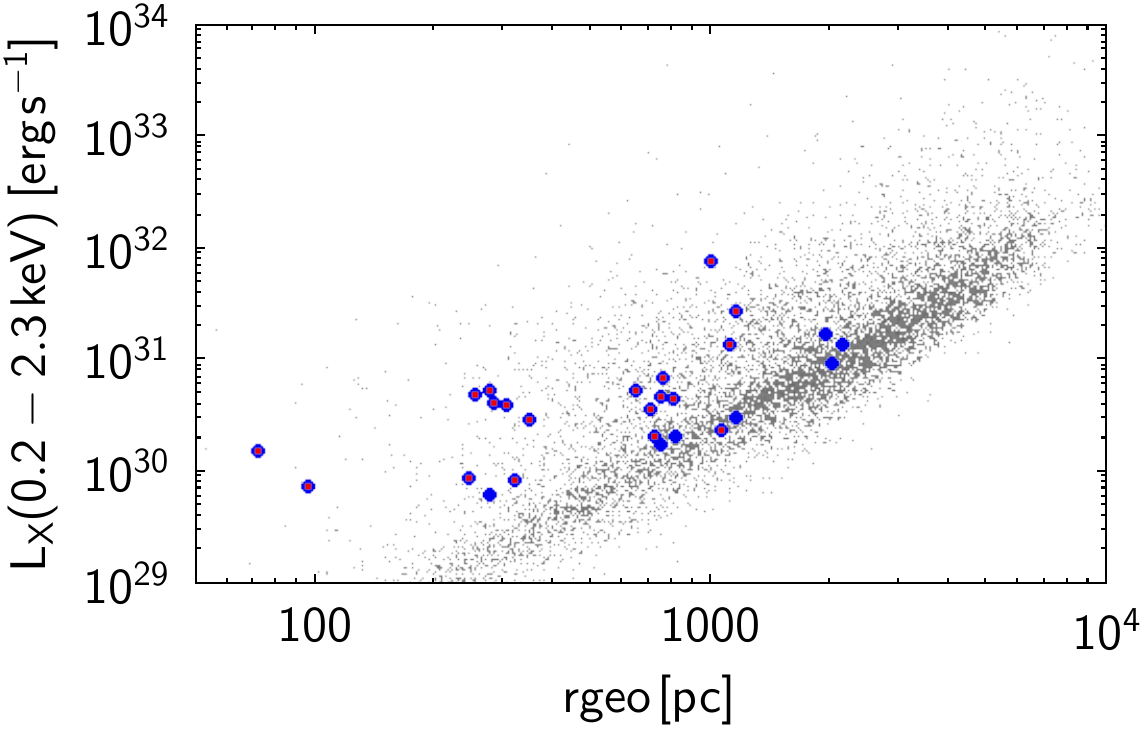}}
\hfill
\resizebox{0.49\hsize}{!}{\includegraphics{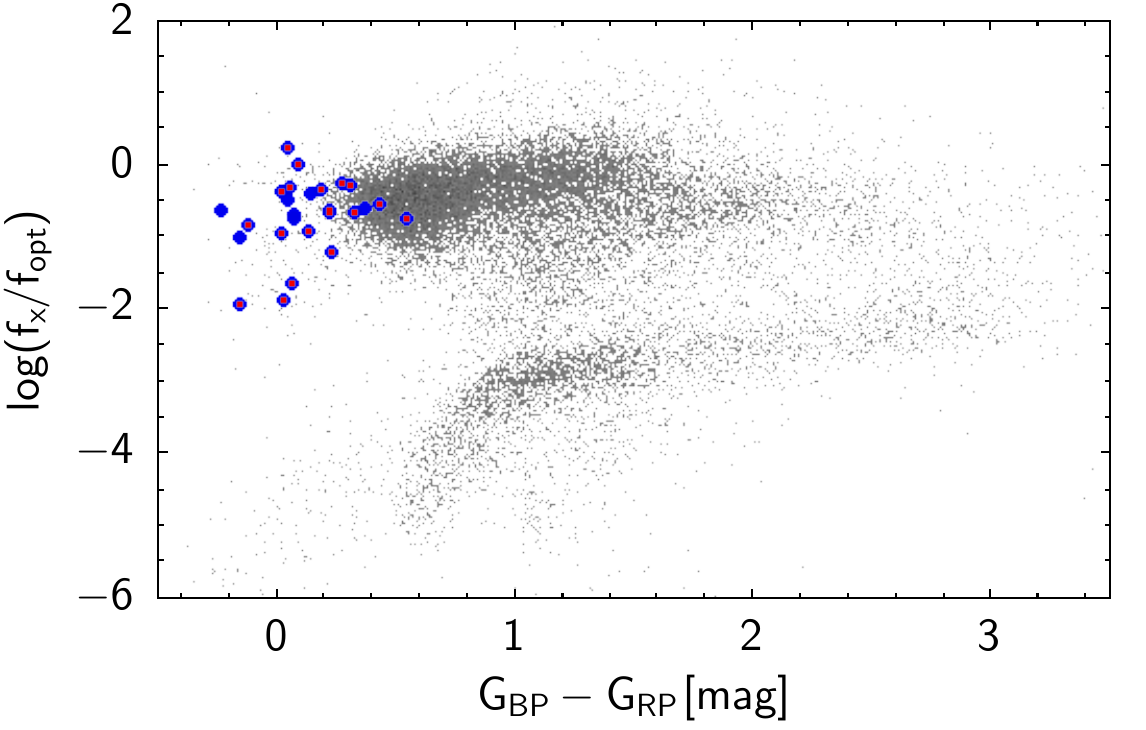}}
\resizebox{0.49\hsize}{!}{\includegraphics{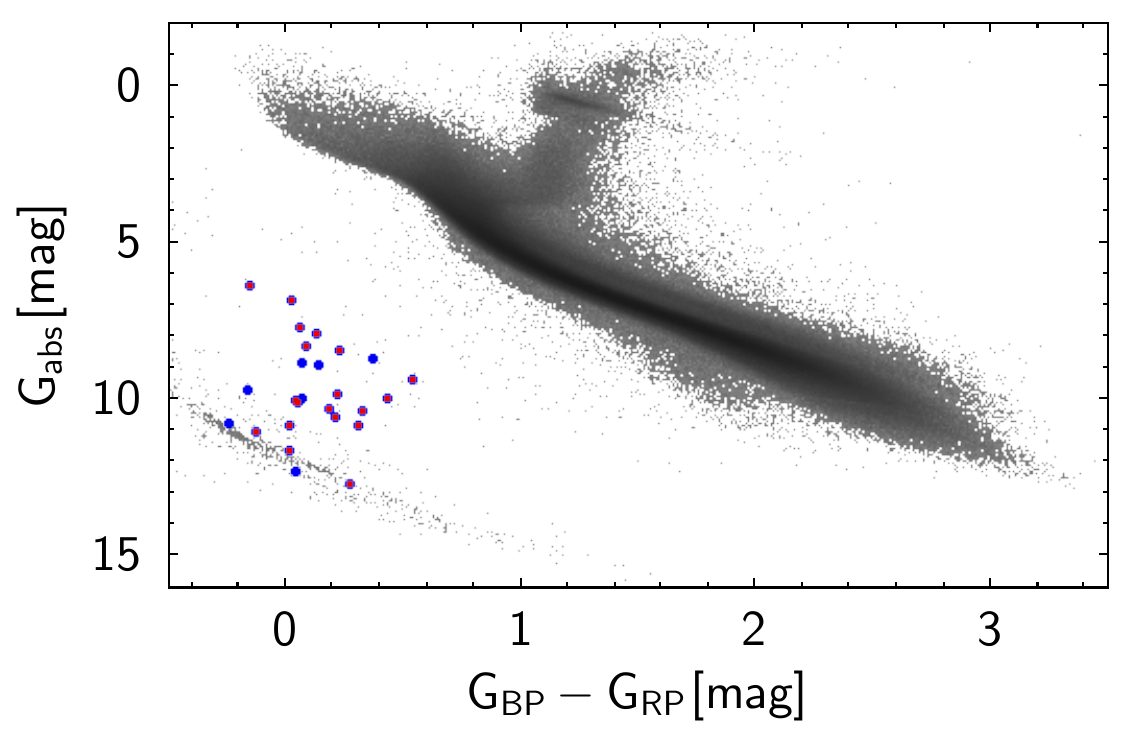}}
\hfill
\resizebox{0.49\hsize}{!}{\includegraphics{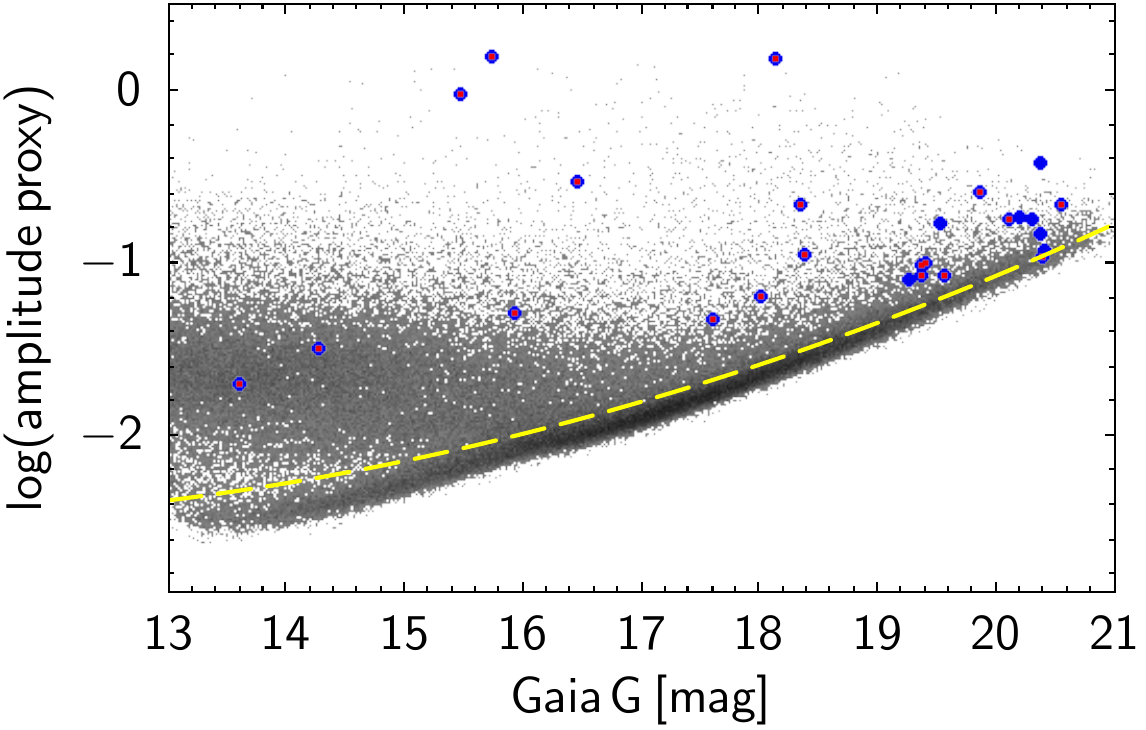}}
\caption{Diagnostic diagrams for AM CVn objects. The color scheme is the same as in Fig.~1.}
\label{f:amcvns}
\end{figure*}

\begin{figure*}
\resizebox{0.49\hsize}{!}{\includegraphics{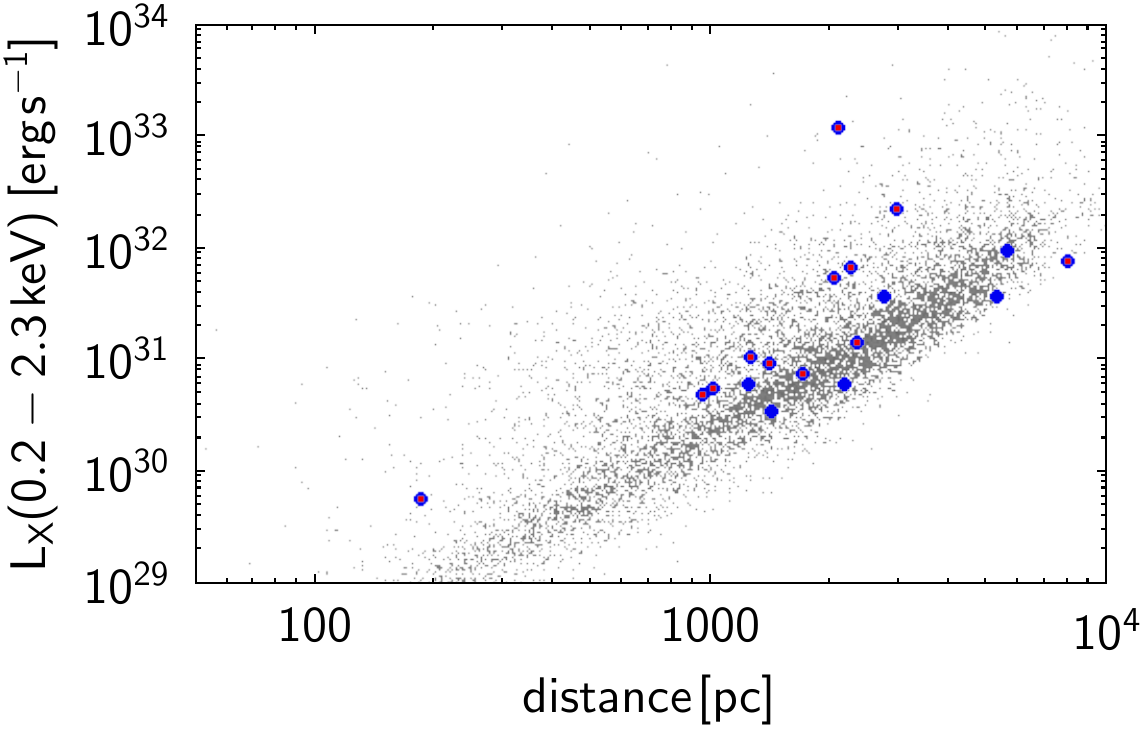}}
\hfill
\resizebox{0.49\hsize}{!}{\includegraphics{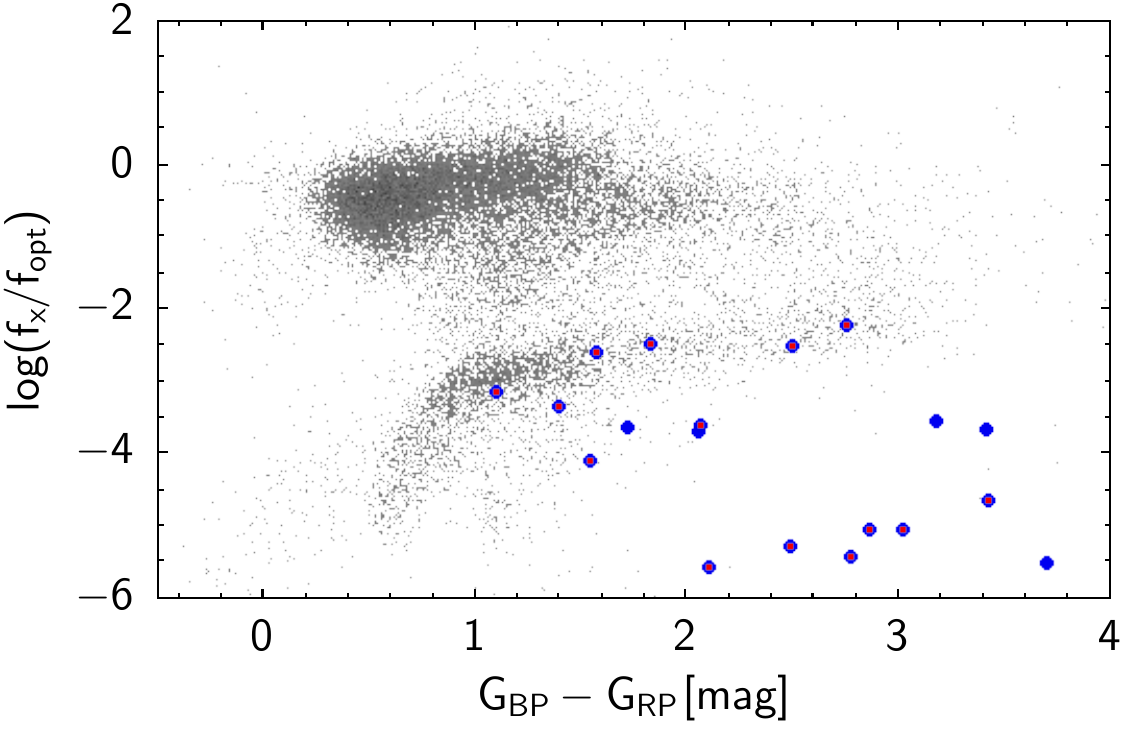}}
\resizebox{0.49\hsize}{!}{\includegraphics{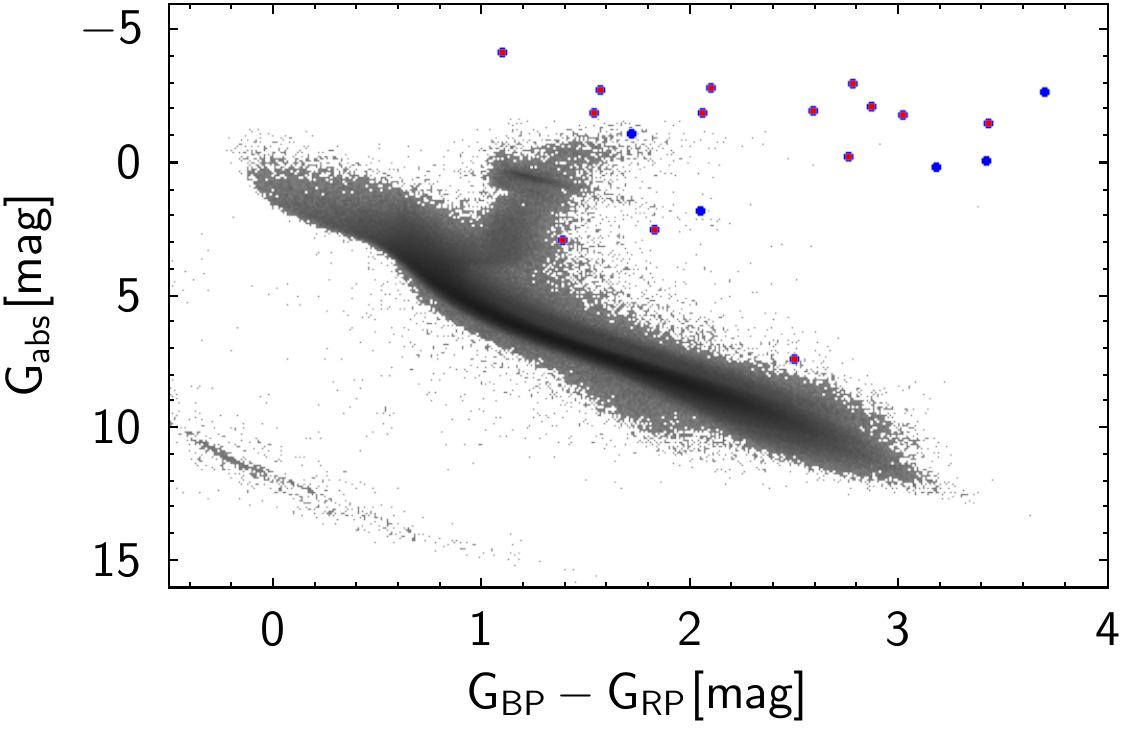}}
\hfill
\resizebox{0.49\hsize}{!}{\includegraphics{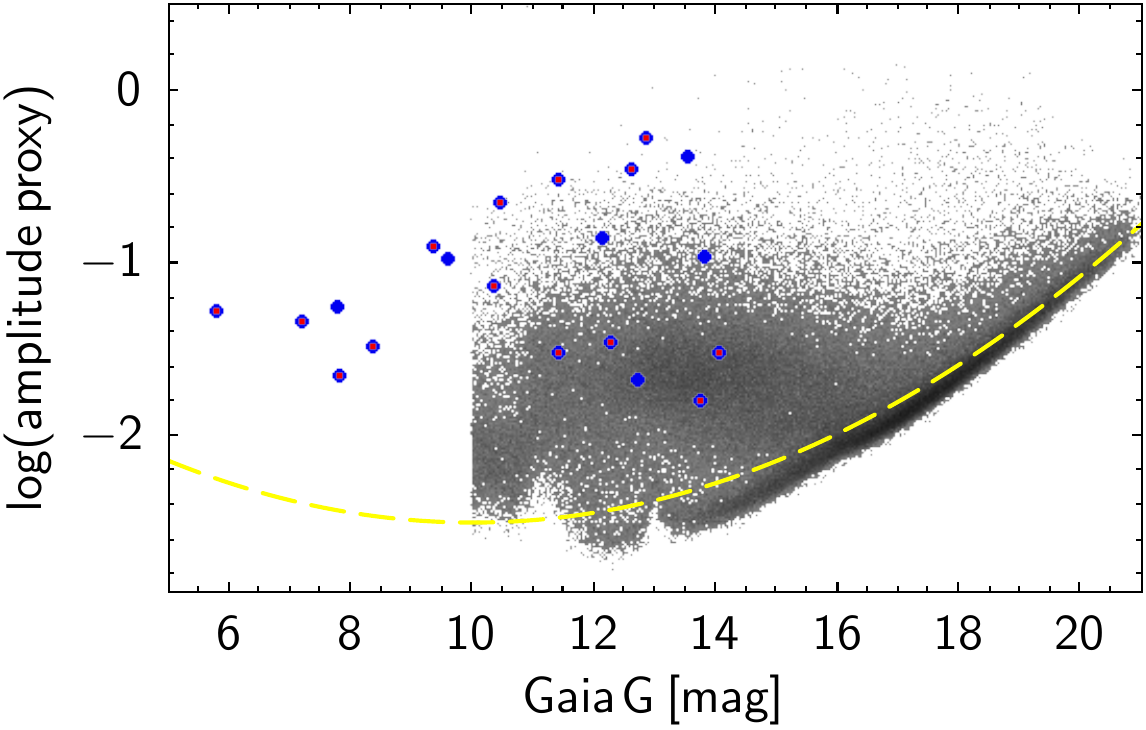}}
\caption{Diagnostic diagrams for SySts. The color scheme is the same as in Fig.~1.}
\label{f:symbs}
\end{figure*}

\subsubsection{Optical variability from \textit{\gai}}
\cite{eyer+20} defined a proxy for the variability amplitude. The logarithm of this quantity is shown in Fig.~\ref{f:var}, which immediately reveals that almost all the known CVs are strongly variable \textit{\gai} sources, that is, they~are located far above the yellow line indicating constant sources. Not unexpectedly, the DN are the most variable class of CVs. The variability amplitude of the DN apparently decreases for faint magnitudes. 
The polars show strong variability at all brightness levels, which is due to strong orbital variability mainly through cyclotron beaming and possibly due to their frequent changes between high and low accretion states. The polars have a few members very close to the nonvariable limit. These are most likely systems encountered by \textit{\gai} in their low accretion states. All of the IPs and the NLs (with very few exceptions) have a medium to high variability amplitude at all brightness states.

\subsection{X-ray spectral analysis of bright polars}
\label{s:polars}
An X-ray spectral analysis with \xspec was performed for all polars with more than 500 counts that were detected in S45 in the energy band $0.2-2.3$\,keV of all seven camera modules. This applies to 22 objects. The data for the source and the background were extracted from the calibrated event lists (pipeline processing version c020) using eSASS \citep{brunner+22}. They were spectrally binned with a minimum of 25 photons per bin to allow $\chi^2$ minimization in spectral fitting. For this exercise only data for TM8\footnote{TM stands for Telescope Module. \ero has seven TMs, TM8 stands for the sum of TM1, TM2, TM3, TM4, and TM6.} were used to avoid problems with the light leaks of TM5 and TM7 \citep{predehl+21}. Following standard practice for X-ray spectral fits of polars, we initially used just a thermal model with fixed temperature corresponding to 15\,keV, modified by absorption in cold interstellar matter ({\tt TBABS*APEC} in \xspec terminology). The chosen temperature was assumed to reliably represent the thermal plasma emission in the observed spectra. The effective area of \ero above 2.3 keV is too small to allow an independent measurement of the expected high(er) temperature. The chosen value is in accord with the value found for the prototypical system AM Her \citep{schwope+20}.

Large residuals were observed for the soft polars with HR1$<0.3$. For these, a second model was applied in which a low-temperature blackbody was added ({\tt TBABS*(BBODY+APEC)}). This model revealed acceptable or even good fits for all the objects. The results of the spectral fitting are summarized in Table~\ref{t:polars}. The first column identifies the object, the second column gives the magnetic field strength in the main accretion region, which is assumed to be responsible for most of the X-ray radiation. A few of the polars show two-pole accretion (e.g.~UZ For and VV Pup), where the second pole has a higher magnetic field but a much lower accretion rate than the main pole. We thus assign the observed flux to the main pole with the lower field strength. The third column lists the number of photons available for a spectral analysis. The further columns give the $\chi^2/d.o.f.$ as a goodness of fit criterion, the best-fitting blackbody temperature, the absorption column density, the fluxes in the soft and the hard components and the flux ratio.

Only half of the polars analyzed here show a soft blackbody-like component. This is contrary to one of the classical hallmarks, namely the existence and prevalence of a soft blackbody-like component, discussed in the literature as the 'famous soft X-ray puzzle of the AM Herculis stars'  \cite[for thorough discussions of observations and theoretical explanations, see][and references therein or references to those papers]{frank+88, beuermann+schwope94, ramsay+cropper04}. Using a larger set of \xmmn spectra of polars already \cite{ramsay+cropper04} have shown that the soft X-ray component is missing in many systems and that the excess is milder than previously assumed. Indeed, many polars found since then lack the soft component, the finding of \cite{ok+24} representing the most recent example. However, there is a number of polars that show a soft component and several of those were among the systems that were analyzed here. Fig.~\ref{f:polsoft} sheds new light on the strength of a soft excess as a function of the magnetic field strength in the main accretion region. Similar figures were presented by \cite{beuermann+schwope94, beuermann+burwitz95} and by \cite{ramsay+cropper04} on the basis of \ros data. While the earlier papers from the 90ies showed a clear trend of the flux ratio as a function of $B$ in the sense that soft X-rays were dominating the hard X-rays (or hard X-rays were suppressed) the later analysis by \cite{ramsay+cropper04} showed no evidence of a trend. These authors concluded that the trend of the flux ratio against magnetic field was largely due to the small sample  used earlier. The figure we show here seems to indicate the old trend: a lower flux ratio $f_{\rm APEC}/f_{\rm BB}$ with larger $B$; however, we also have a small sample size if only polars are considered. While the question of whether or not there is a trend is not finally settled with the sample of polars analyzed here, the other important number that is derived is the blackbody temperature, which ranges from about 20 eV (BL Hyi) to almost 60 eV for 1RXS J100734.4-201731. This value is in agreement with a high state \xmmn observation that revealed 59 eV \citep{thomas+12}. Interestingly, this high value of $kT_{\rm BB}$ was found for the object with the highest field strength of 94 MG. 

The very low blackbody temperatures of the polars make the flux determination and hence bolometric corrections uncertain. The picture becomes more complete, if the soft IPs (see the next section) are included. As is shown there, none of the IPs (with the possible exception of PQ Gem) is dominated by soft X-rays. We assign the same assumed field strength of 5 MG to all the soft IPs, no direct measurement of it was ever possible but optical polarization was found for some of the IPs. The inclusion of the IPs suggests that hard X-rays are suppressed efficiently for accretion columns with $B\gtrapprox 20$\,MG.

\begin{table*}[t]
\centering
\caption{X-ray spectral parameters for the brightest polars in S45. Here, we provide the magnetic field in the prime pole, counts in TM8, fitted blackbody temperatures and hydrogen column densities, the value of the fit statistics, and the implied bolometric fluxes for the soft blackbody and the hard plasma component. The temperature in the {\tt APEC} model was fixed to 15 keV. Errors were derived for a 90\% confidence interval. }
\label{t:polars}
\begin{tabular}{lrrcccccc}   
\hline\hline
Name &  $B$ & counts & $T_{bb}$ & $n_{\rm H}$ & $\chi^{2}/dof$ & $F_{\text{soft}}$ & $F_{\text{hard}}$ & $F_{\text{soft}}/F_{\text{hard}}$\\
&  & & [eV] & [$10^{20}$cm$^{-2}$] & & & & \\   \hline
1RXS J100734.4-201731 & 94 &  377 & $57.0^{+2.7}_{-3.4}$ & $0.0^{+0.7}_{-0.0}$ & 1.01 & $0.62^{+0.13}_{-0.13}$ & $0.05^{+0.01}_{-0.04}$ & $11.7^{+35.1}_{-4.1}$ \\   
1RXS J152506.9-032647 & -- &  618 & -                    & $1.8^{+3.1}_{-1.8}$ & 0.83 & --                     & $0.65^{+0.07}_{-0.07}$ & -- \\   
BL Hyi                & 23 & 1288 & $18.7^{+5.5}_{-4.3}$ & $0.0^{+82}_{-0.0}$  & 1.06 & $23^{+320}_{-20}$      & $0.93^{+0.01}_{-0.09}$ & $25^{+385}_{-21}$ \\   
CD Ind                & 11 &  825 & -                    & $1.5^{+1.7}_{-1.5}$ & 1.30 & --                     & $1.26^{+0.11}_{-0.10}$ & -- \\
CTCV J1928-5001       & 20 &  786 & $12.2^{+0.5}_{-1.1}$ & $0.0^{+1.6}_{-0.0}$ & 1.03 & n.a.                   & n.a.& n.a.\\
CW Hyi                & 13 &  968 & $51.6^{+5.0}_{-8.9}$ & $0.3^{+1.9}_{-0.3}$ & 0.86 & $0.23^{+0.46}_{-0.07}$ & $0.46^{+0.04}_{-0.04}$ & $0.5^{+1.1}_{-0.2}$ \\
eRASS J1425-31        & -- &  375 & -                    & $6.2^{+5.4}_{-4.7}$ & 1.13 & --                     & $0.53^{+0.08}_{-0.08}$ & -- \\ 
IGR J14536-5522       & 20 &  830 & -                    & $5.2^{+2.4}_{-2.2}$ & 1.17 & --                     & $0.97^{+0.09}_{-0.09}$ & --\\
IW Eri                & -- &  704 & -                    & $0.0^{+1.5}_{-0.0}$ & 0.74 & --                     & $0.59^{+0.05}_{-0.03}$ & --\\ 
PBC J0658.0-1746      & -- &  402 & $32^{+30}_{-17}$     & $0.0^{+3.0}_{-0.0}$ & 0.78 & $0.2^{+2.3}_{-0.1}$    & $0.70^{+0.08}_{-0.08}$ & $0.3^{+3.7}_{-0.2}$\\   
PBC J0706.7+0327      & -- &  798 & -                    & $0.0^{+1.0}_{-0.0}$ & 0.76 & --                     & $1.7^{+0.1}_{-0.1}$    & --\\
RX J0154.0-5947       & -- &  721 & $38.1^{+5.9}_{-5.6}$ & $0.0^{+82}_{-0.0}$  & 1.54 & $0.63^{+0.56}_{-0.26}$ & $0.45^{+0.04}_{-0.04}$ & $1.4^{+1.5}_{-0.6}$\\
RX J0649.8-0737       & -- &  482 & $48.4^{+9.5}_{-7.8}$ & $4.0^{+3.2}_{-2.7}$ & 1.04 & $1.9^{+5.7}_{-1.3}$    & $0.43^{+0.06}_{-0.08}$ & $4^{+17}_{-3}$\\   
Swift J0503.7-2819    & -- &  548 & -                    & $3.6^{+3.2}_{-2.8}$ & 1.10 & --                     & $0.45^{+0.06}_{-0.05}$ & --\\   
UW Pic                & -- &  647 & $35.4^{+3.5}_{-3.5}$ & $0.0^{+82}_{-0.0}$  & 1.13 & $0.47^{+0.22}_{-0.14}$ & $0.14^{+0.02}_{-0.02}$ & $3.3^{+2.2}_{-1.2}$ \\  
UZ For                & 53 &  396 & $32.7^{+2.5}_{-4.8}$ & $0.0^{+1.7}_{-0.0}$ & 0.35 & $1.8^{+0.9}_{-0.6}$    & $0.08^{+0.01}_{-0.03}$ & $22^{+27}_{-9}$\\      
V1033 Cen             & 20 & 2967 & -                    & $14.6^{+1.6}_{-1.5}$& 1.12 & --                     &$2.7^{+0.1}_{-0.1}$     & -- \\   
V1043 Cen             & 56 &  532 & $36.9^{+3.0}_{-5.7}$ & $0.3^{+2.0}_{-0.3}$ & 0.77 & $2.1^{+7.7}_{-0.64}$   & $0.07^{+0.02}_{-0.02}$ & $32^{+196}_{-15}$\\   
V1309 Ori             & 61 &  647 & $43.2^{+4.4}_{-4.3}$ & $4.4^{+1.8}_{-1.5}$ & 1.01 & $5.4^{+7.4}_{-2.7}$    & $0.06^{+0.03}_{0.03}$  & $85^{+282}_{-56}$\\ 
V347 Pav              & -- &  377 & -                    & $0.0^{+1.8}_{-0.0}$ & 1.12 & --                     &$0.38^{+0.04}_{-0.4}$   & -- \\ 
V834 Cen              & 23 &  755 & $33.6^{+3.0}_{-3.0}$ & $0.0^{+82}_{-0.0}$  & 1.27 & $3.3^{+5.1}_{-0.5}$    & $0.72^{+0.06}_{-0.08}$ & $4.6^{+8.6}_{-1.0}$\\   
VV Pup                & 31 &  494 & $30.9^{+4.3}_{-4.6}$ & $0.0^{+0.8}_{-0.0}$ & 0.63 & $2.9^{+4.8}_{-0.5}$    & $0.66^{+0.07}_{-0.07}$ & $4.4^{+8.7}_{-1.2}$\\   \hline
\end{tabular} 

\end{table*}

\begin{table*}
\centering
\caption{Spectral fits for the seven soft IPs detected in S45. Here, we provide the object names and the numbers of photons registered in S45 followed by the spectral parameters. The bolometric fluxes are given in $10^{-11}$ erg cm$^{-2}$ s$^{-1}$, the hydrogen column density $N_{\rm H}$ is given in units of $10^{20}$\,cm$^{-2}$. Errors given refer to 90\% confidence.}
\label{t:ips}
\begin{tabular}{lrcccccccc}
\hline\hline
Name     & counts & $N_{\rm H}$ & $kT_{\rm bb}$[eV] & $kT_{\rm lit}$[eV] & $F_{\rm bol,bb}$ & $F_{\rm bol,apec}$ & $\chi^2$  \\ \hline
PQ Gem   & 3257   & $0.2^{+0.009}_{-0.002}$ & $47^{+4}_{-5}$   & $47.6^{+2.9}_{-2.4}$ & 2.8  & 2.0  & 1.23 \\
V418 Gem &  442   & $<+5$       & $84^{+14}_{-18}$ & $84 - 88$ & 0.14 & 0.2  & 0.51 \\
UU Col   &  598   & $<1$        & $65^{+3}_{-1}$   & $73^{+20}_{-9}$ & 0.02 & 0.28 & 0.99 \\
V667 Pup & 1503   & $9^{+4}_{-4}$    & $74^{+9}_{-8}$   & -- & 1.0  & 1.3  & 0.74 \\
BG CMi   &  479   & $6^{+10}_{-6}$   & $60^{+30}_{-20}$ & -- & 0.9  & 13.4 & 1.11 \\
TX Col   & 3248   & $3^{+2}_{-2}$    & $95^{+14}_{-12}$ & -- & 0.14 & 1.3  & 1.14 \\
NY Lup   & 8705   & $9^{+1}_{-1}$    & $96^{+4}_{-2}$   & $104^{+21}_{-23}$ & 1.9  & 3.6  & 1.10 \\ \hline
\end{tabular}
\end{table*}

\begin{table*}[t]  
\centering
\caption{SySts as listed by \cite{akras+19} detected in S45. Given are the object names, their \textit{Gaia} ID, the number of photons detected in S45, the \ero X-ray coordinates, the separation between the \textit{\gai} and the \ero coordinates in arcsec and normalized to 1$\sigma$ of the X-ray positional uncertainty, the distance (in pc), and information pertaining to whether the object was known before as an X-ray source and, if available, which subtype was assigned}
\label{t:symbs}
\begin{tabular}{lcrrrrrrc}
\hline\hline
Name & \textit{Gaia} ID & \# cts & RA & DEC & Sep. & NSep. & Dist & X-det  \\ \hline
StH$\alpha$ 32 & 3229441606998725888 & $233$ & $69.4403$ & $-1.3200$ & 0.64 & 0.45 & 10221 & $\alpha$  \\
V1261 Ori& 3014902188065355264 & $192$ & $80.5777$ & $-8.6657$ & 1.80 & 1.11 & 375 & yes \\
GH Gem & 3160625132721733888 & $17$ & $106.0523$ & $12.0583$ & 5.37 & 1.44 & 963 & -- \\
ZZ CMi & 3155368612444708096 & $12$ & $111.0524$ & $8.8961$ & 21.85 & 3.07 & 1245 & yes\\
NQ Gem & 868424696282795392 & $16$ & $112.9751$ & $24.5042$ & 6.95 & 1.39 & 1012 &  $\beta/\delta$ \\
WRAY 15-157& 5597659331053579008 & $13$ & $121.6442$ & $-28.5335$ & 3.49 & 0.77 & 5655 & --  \\
RX Pup & 5533427747232584192 & $25$ & $123.5512$ & $-41.7080$ & 0.37 & 0.10 & 2744 & $\beta$  \\
SS73 17 & 5258810806805860992 & $47$ & $152.7625$ & $-57.8038$ & 0.66 & 0.27 & 1261 &  yes \\
Hen 3-461 & 5360768073375283712 & $12$ & $159.7845$ & $-51.4035$ & 3.54 & 0.65 & 1431 & $\delta$  \\
SY Mus & 5237239075896985728 & $28$ & $173.0427$ & $-65.4205$ & 2.72 & 0.96 & 1723 &  -- \\
BI Cru & 6054612927609974656 & $200$ & $185.8580$ & $-62.6383$ & 1.81 & 1.26 & 2958 & $\beta/\delta$ \\
RT Cru & 5861308338109134208 & $24$ & $188.7245$ & $-64.5662$ & 2.44 & 0.67 & 2347 & $\delta$ \\
Hen 2-87 & 5862972414572104192 & $11$ & $191.4472$ & $-63.0099$ & 2.31 & 0.46 & 5301 & yes \\
SS73 38& 5861762230260087936 & $9$ & $192.8575$ & $-64.9987$ & 3.51 & 0.58 & 2186 & -- \\
V840 Cen & 6063900742825792000 & $89$ & $200.2071$ & $-55.8379$ & 2.73 & 1.39 & 2065 & -- \\
AE Cir & 5799841957262188928 & $9$ & $221.2177$ & $-69.3920$ & 6.14 & 1.19 & 8006 &  -- \\
IGR J16194-2810${^1)}$ & 6039227323926220544 & $1110$ & $244.8892$ & $-28.1275$ & 1.36 & 1.33 & 2101 &  $\gamma$ \\
SWIFTJ171951.7-300206${^2})$ & 4059196651237110272 & $67$ & $259.9663$ & $-30.0329$ & 2.63 & 1.10 & 187 & yes\\
RR Tel & 6448785024330499456 & $115$ & $301.0780$ & $-55.7261$ & 1.72 & 0.93 & 13162 &  $\alpha$ \\
Hen 4-137 & 6069918816627209856 & $22$ & $203.9887$ & $-51.2037$ & 2.91 & 0.81 & 1414 &  -- \\ \hline
\end{tabular}

Note 1) IGR J16194-2810 is considered being a Symbiotic X-ray binary, not a WD accreting object \citep[e.g.][]{bozzo+24} \\

Note 2) SWIFTJ171951.7-300206 is reclassified here as a likely X-ray active nearby star.

\end{table*}

\begin{table}[]
\centering
\caption{Comparison of X-ray fluxes of SySts observed with \xmmn and \ero. For \xmmn, the summed catalog fluxes from the 4XMM bands 1, 2, and 3 were used ($0.2-2.0$\,keV) and scaled to the \ero band ($0.2-2.3$\,keV). Flux units are $10^{-14}$\,\fergs. The last column lists the flux ratio $f_{\rm eRO}/f_{\rm XMM}$. }
\label{t:sysfrat}
\begin{tabular}{lr@{$\pm$}lr@{$\pm$}lr@{$\pm$}l}
\hline\hline
Name& \multicolumn{2}{c}{Flux$_{\rm XMM}$}  & \multicolumn{2}{c}{Flux$_{\rm eRO}$} & \multicolumn{2}{c}{Ratio} \\
\hline
V1261 Ori & 114.2  & 0.6  & 33.2 & 2.6 &  0.29 & 0.02\\
NQ Gem    &  14.71 & 0.24 &  4.5 & 1.3 &  0.31 & 0.09\\
RX Pup    &  34.82 & 0.60 &  4.1 & 1.0 &  0.12 & 0.03\\
Hen 3-461 &   0.14 & 0.03 &  1.4 & 0.5 &  9.7  & 4.1 \\
RT Cru    &   4.4  & 0.1  &  2.2 & 0.5 &  0.50 & 0.11\\
Hen 2-87  &   1.3  & 0.4  &  1.1 & 0.4 &  0.83 & 0.41\\
RR Tel    &  24.5  & 0.9  &  27  & 3   &  1.10 & 0.13\\
\hline
\end{tabular}
\end{table}

\begin{figure*}
\resizebox{0.49\hsize}{!}{\includegraphics{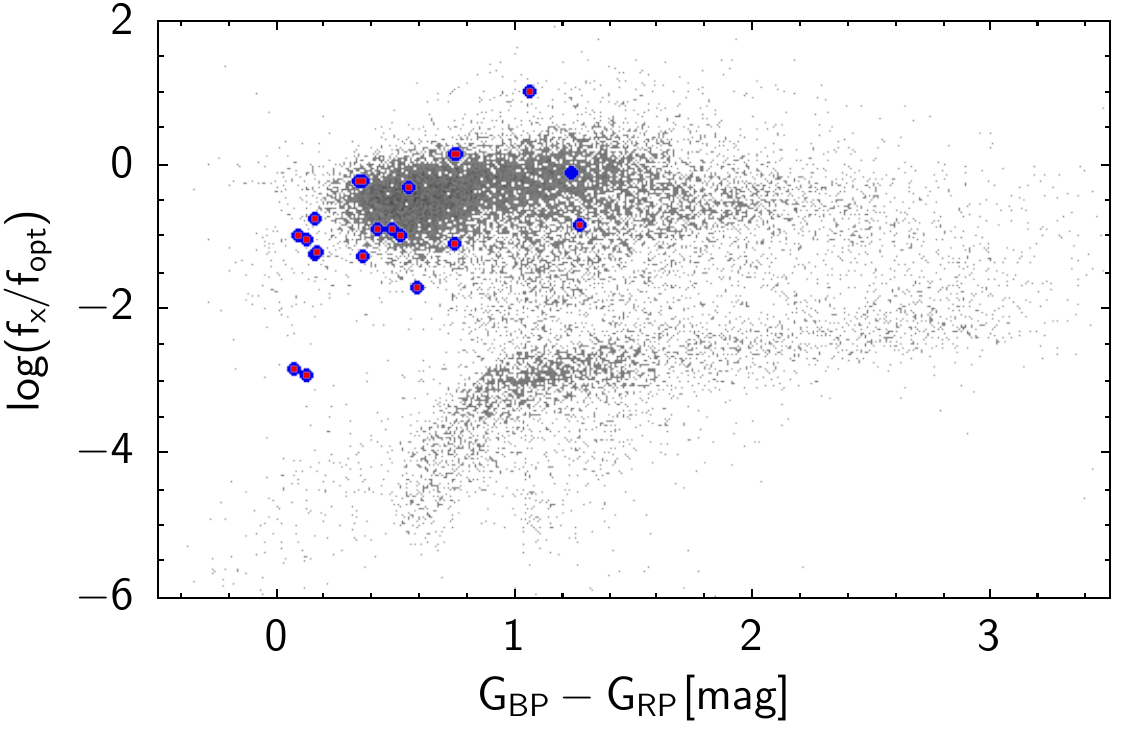}}
\hfill
\resizebox{0.49\hsize}{!}{\includegraphics{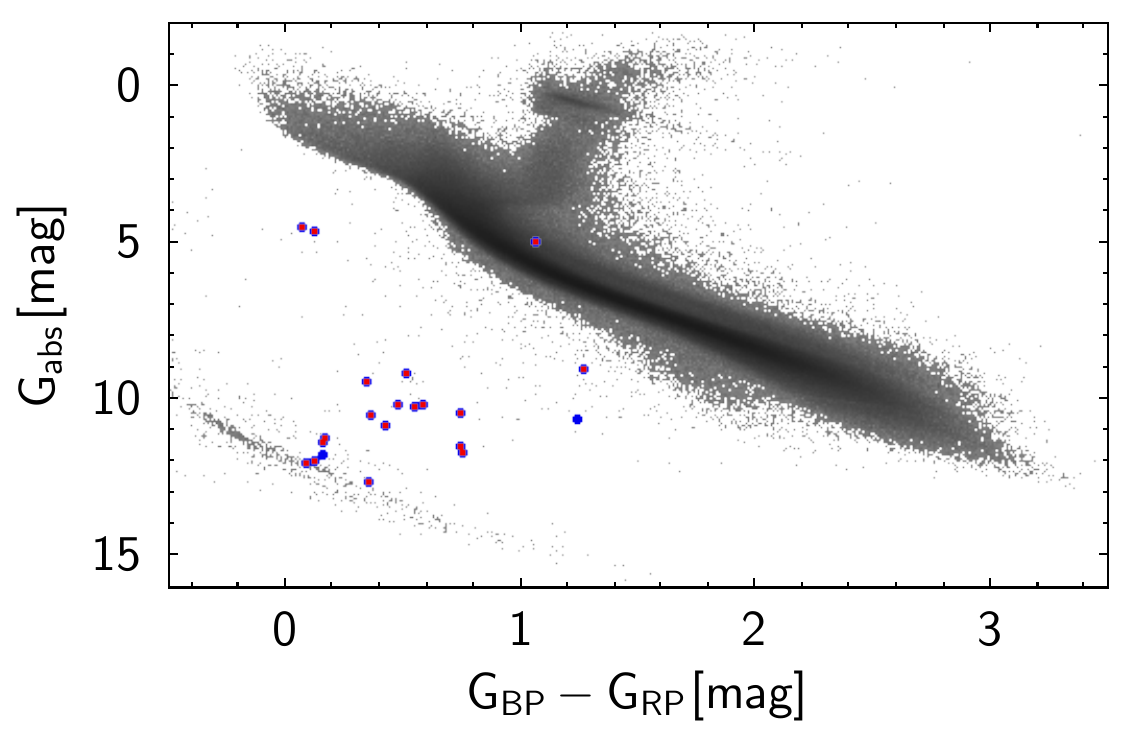}}
\resizebox{0.49\hsize}{!}{\includegraphics{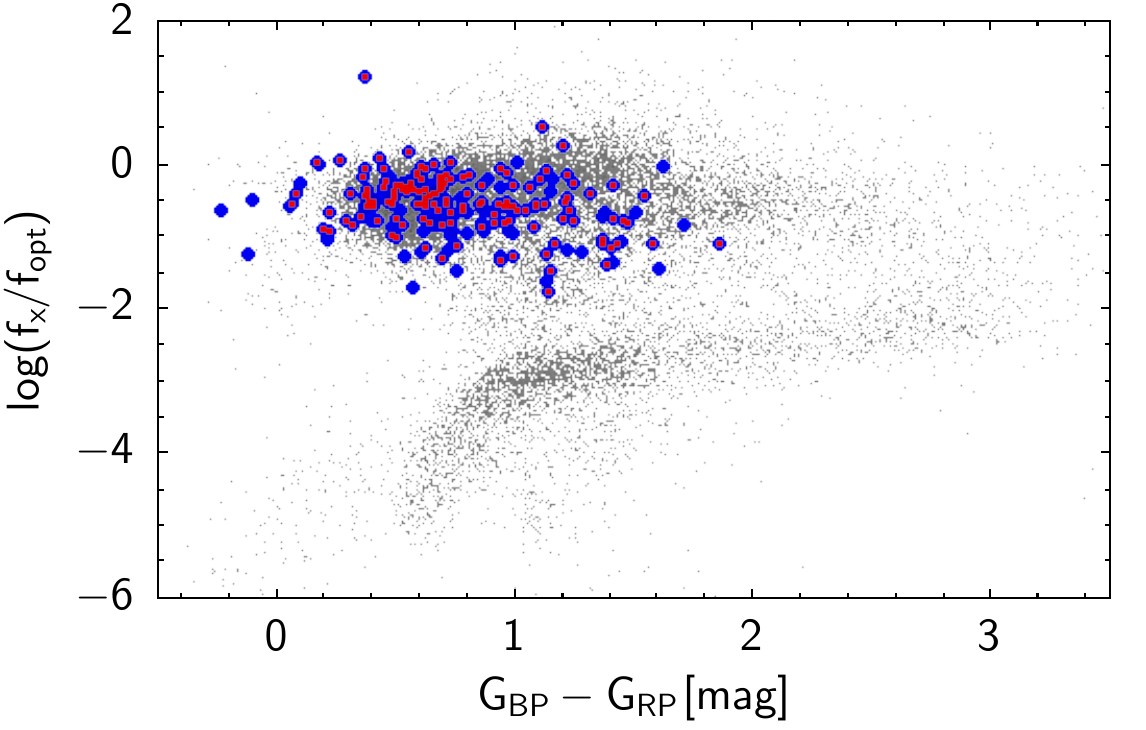}}
\hfill
\resizebox{0.49\hsize}{!}{\includegraphics{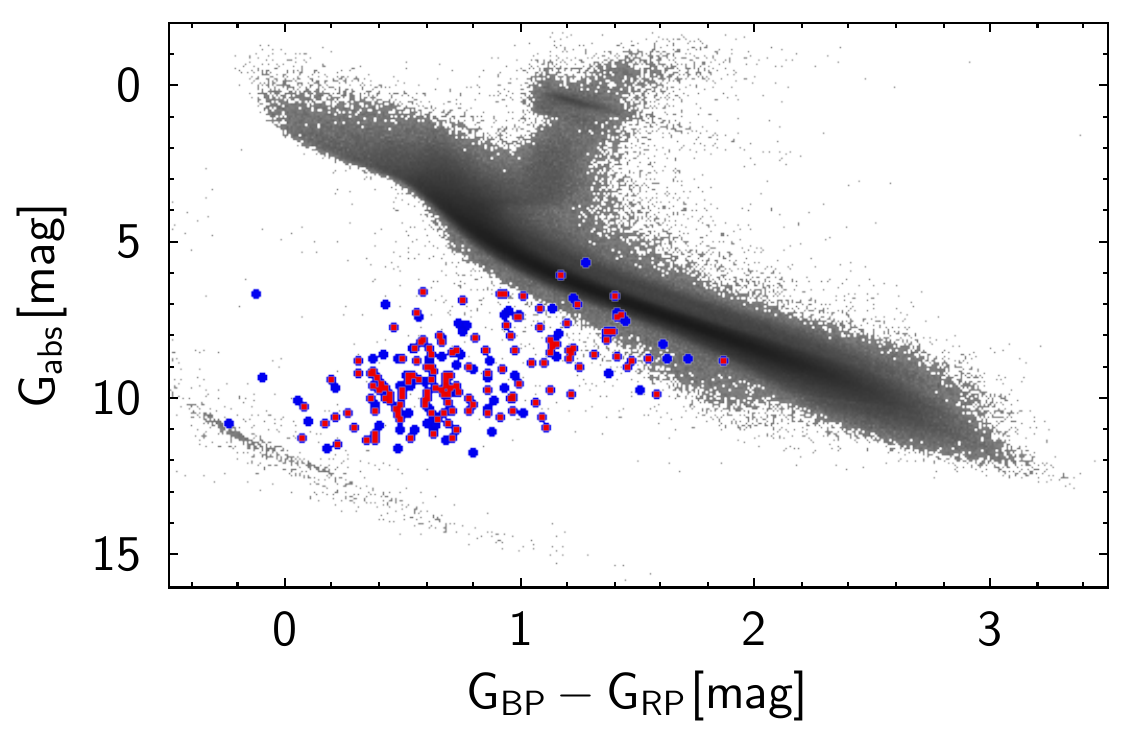}}
\resizebox{0.49\hsize}{!}{\includegraphics{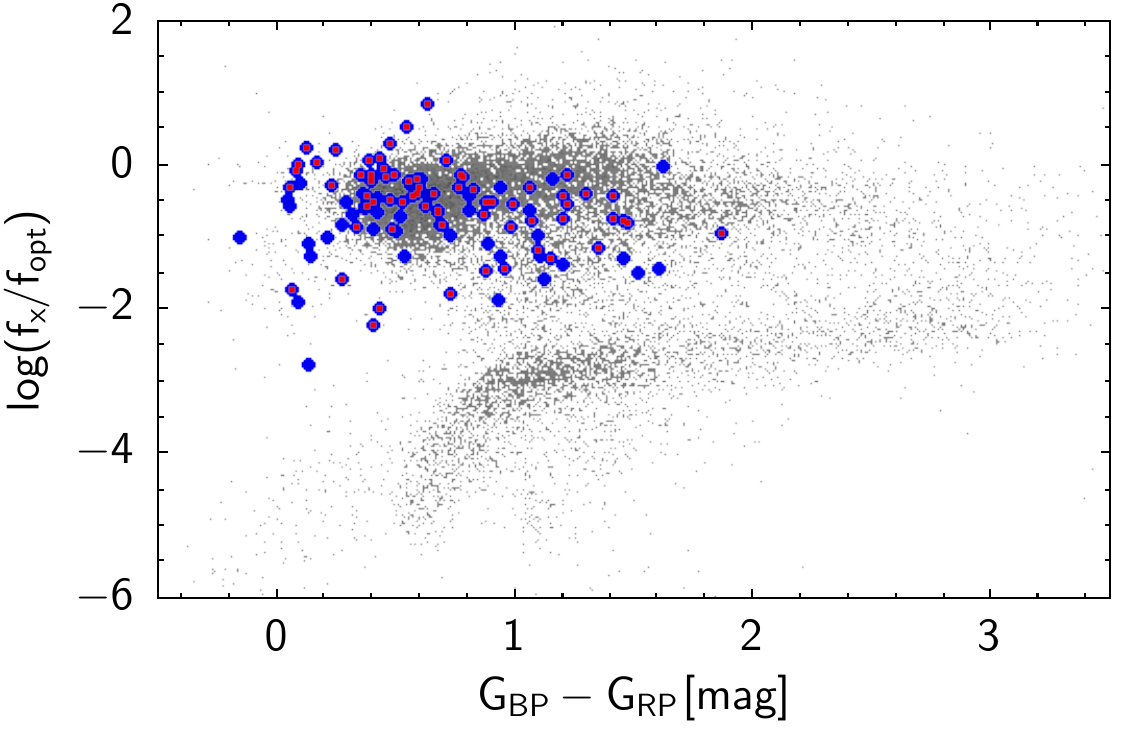}}
\hfill
\resizebox{0.49\hsize}{!}{\includegraphics{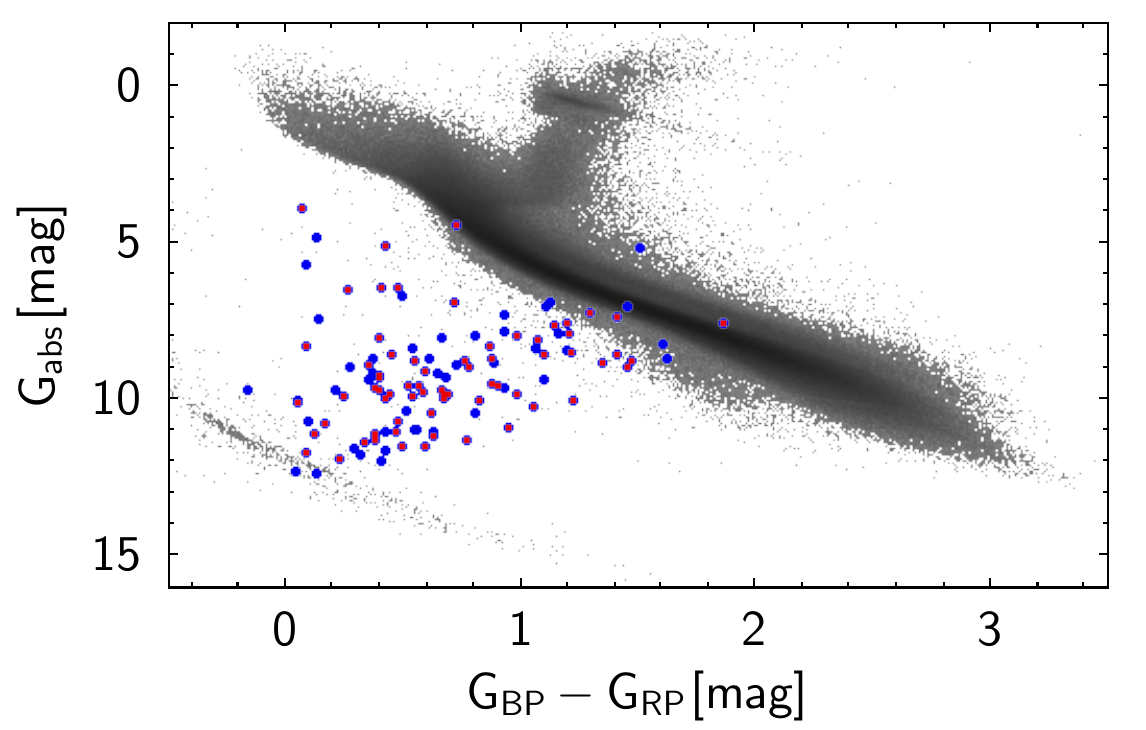}}
\caption{CCDs (left) and CMDs (right) for the mixed samples from \cite{pala+20}, the CRTS (middle row), and the SDSS (bottom row).  The color scheme for CVs is the same as in Fig.~1.}
\label{f:mixed}
\end{figure*}

\subsection{X-ray spectral analysis of bright, soft intermediate polars}
\label{s:softips}
The X-ray spectra of the seven candidate soft IPs were analyzed in a similar manner as the polars (for the technical part the reader is referred to the previous section). Initially, also just an absorbed thermal model was used with a fixed temperature of 15\,keV. Using a higher $T$ has a negligible effect on the quality of spectral fits but will lead to higher fluxes. The chosen $T$ is likely a lower limit for the hard IPs and the bolometric correction to be applied is therefore uncertain. 

As expected from the hardness ratio HR1, the initial model gave in no case a satisfactory fit to the data. Large residuals were found at soft X-ray energies in the first place. We then added a blackbody component, which improved the fit quality in all cases, thus proving that the initial selection based on hardness ratios was successful. The fits including a soft blackbody component often left residuals at high energies indicating even harder spectra. Compton reflection parameterized with the {\tt reflect} model revealed the needed spectral hardening. We thus adopted finally a model with the combination of blackbody emission and a thermal optically thin plasma component modified by reflection, i.e.{\tt TBABS*(BBODY+reflect*APEC)}, which in all but one (PQ Gem) cases revealed good fits. The results of the spectral analysis are summarized in Table~\ref{t:ips}. The amount of absorption is in almost all cases negligible. The fluxes given in Table~\ref{t:ips} are bolometric fluxes for the 15 keV spectrum used in the fits. The bolometric correction for a 40\,keV spectrum is a factor of 1.48 larger. Taking the values in the Table at face value the flux ratio between the hard plasma and the soft blackbody components varies between $F_{\rm APEC}/F_{\rm BB} = 0.7$ and 14. There is only one soft-dominated IP, PQ Gem. However, if the bolometric correction for 40 keV is used, even PQ Gem has a balanced energy output. Finally, we note that for the three IPs TX Col, BG CMi, and V667 Pup a soft component is reported and the blackbody temperature measured for the first time.

A word of caution seems to be in place regarding the robustness of the parameters of the emission components documented in Table \ref{t:ips}. These are based on the assumption that the observed emission can be described by the sum of two thermal emission components modified by reflection and cold interstellar absorption. It is known, however, that more complex absorbers (intrinsic partial covering and ionized absorbers) and more complex emission models (multi-temperature shocked plasma, photoionized plasma radiation) might play their role in shaping the observed spectra \citep[see e.g.][and references therein]{islam+mukai21}. The inclusion of such more complex models would have an effect on the parameter values as derived here for the simpler models, which were chosen to yield adequate representations of the data. However, we regard the presence of an \ero-detected soft component in the seven listed IPs a robust result.

\subsection{AM CVn binaries} 
\label{s:amcvns}
Diagnostic diagrams for the AM CVn systems are shown in Fig.~\ref{f:amcvns}. Nineteen out of 29 in the western galactic hemisphere with distances from \cite{bailer-jones+21} were detected in DR1, and 26 out of 29 (90\%) in S45. Their median X-ray luminosity is low, the lowest among all the classified CV subsamples but still higher than that of the mixed local sample of \cite{pala+20}. On average they are clearly the bluest of all samples characterized in this paper, only a few NLs with their hot disks are similarly blue. They typically have a high X-ray to optical flux ratio, $-0.65$, but three are separate at $<-1.6$ (HP Lib, CR Boo, TIC 378898110), i.e.~they are optically brighter or X-ray fainter than the average AM CVn system. Of these, only CR Boo is outbursting, which, if caught in an outburst optically, could explain the low flux ratio. Three objects show an unusual high proxy of the variability amplitude, namely V803 Cen, CR Boo, and ASASSN-14cc, all three are known to show outbursts and it appears likely that \textit{\gai} observed both during outbursts and quiescence to imply the large amplitude proxy. Their median absolute magnitude is 10.0, as low as that of the DN and the polars. Five of the 19 lie on the WD track in the CMD.

\begin{figure}
\resizebox{\hsize}{!}{\includegraphics{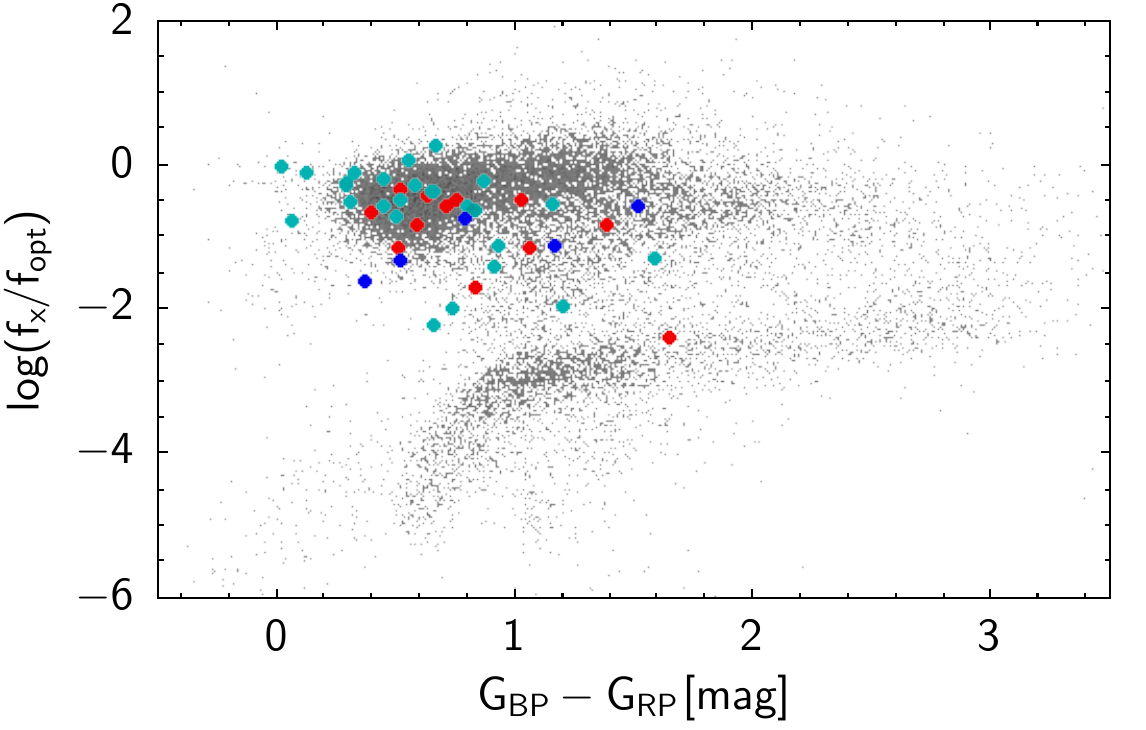}}
\caption{CCD of standard CVs not detected in S45. The flux ratios were computed for upper limit X-ray fluxes. Dwarf novae are indicated with cyan symbols, NLs with blue, and polars with red symbols. No secure IP (4 and 5 stars) was missing at the sensitivity of S45.}
\label{f:uls}
\end{figure}

\subsection{Symbiotic binaries} 
\label{s:systs}
Compared to the standard CVs and the DDs the SySts have a completely different appearance in the diagnostic diagrams and also their mean parameters (see Fig.~\ref{f:symbs} and Tables~\ref{t:matches} and \ref{t:parms}). Of all 400 (315 galactic) sources in Akras' catalog, 60 (56) were listed as known X-ray sources in their catalog.

Of the 144 in the western Galactic hemisphere with \textit{\gai} information only 15 (20) are matching with \ero DR1 (S45) sources, representing a little more than 10\%. The gain in detected SySts in S45 compared to DR1 is rather moderate and most of the SySts were discovered at the detection limit indicating a large hidden population of X-ray emitting SySts in even deeper survey data. In Table~\ref{t:symbs} we list the objects that were detected in S45. The table also lists previous X-ray detections and their likely classification as $\alpha$, $\beta$, $\gamma$ or $\delta$-type using the scheme originally invented by \cite{muerset+97} and further elaborated by \cite{luna+13}.
Seven out of the 20 are new X-ray detections. 

SySts represent the most distant sample $ \langle rgeo\rangle \simeq 2100$\,pc, and, given the nature of the donors, the optically most luminous sample, $\langle G_{\rm abs} \rangle = -1.6$. They appear to be slightly more X-ray luminous than the RLOF standard CVs, which together means that their X-ray to optical flux ratio is at the extreme low end $ \langle \log f_{\rm X}/f_{\rm opt}\rangle = -3.68$. 

In the CMD all but one object are in the M giant region. The one exception, SWIFT J171951.7-300206, lies only 2 mag above the main sequence and is clearly separate from the other objects, which lie $\sim$10 mag above the main sequence. This object was originally described as a SySts by \cite{masetti+12} who obtained a low-resolution identification spectrum. They found a spectral type of M1-2 III and, based on that, estimated a distance of $\sim$16.6\,kpc or $\sim$6.3\,kpc when considering and correcting for interstellar absorption. Using \textit{\gai}, we find a distance of $187.2\pm 0.7$\,pc, which gives an absolute magnitude of 7.37 and an X-ray luminosity of only $\log (L_{\rm X}/\mbox{erg\,s}^{-1})= 29.7$\,\lx. The classification as a Symbiotic star can be revised with high confidence. It appears as if the source would be just a very active late-type main sequence star. The association with the $\gamma$-ray source IGR J17197-3010 suggested by \cite{masetti+12} was rejected already by \cite{luna+12} and \cite{luna+13}, and hence poses no further a complication for its  classification.

The location of SySts in the CMD, the CCD, and the derived X-ray luminosities are much more affected by the level of absorption, both interstellar and intrinsic to the systems, than in any other source class studied here \citep[see e.g.~][]{luna+13}. We made some attempts to correct the absorption-affected parameters but these remained unsuccessful or indicative at most due to either unreliable extinction coefficients $A_0$ from \textit{\gai}\ or too few SySts that were included in the StarHorse catalog compiled by \cite{anders+21}. Hence, absolute magnitudes, colors, X-ray luminosities, and X-ray to optical flux ratios as given in Table~\ref{t:uniqdr1} shall be used with caution. In principle, each of the SySts need an independent detailed study to find reliable intrinsic parameters.

All of the SySts seem to be optically very variable. This is certainly true for objects fainter than tenth magnitude. For brighter objects neither we nor \cite{eyer+20} did fully characterize \textit{\gai} variability and the amplitude proxy, but pronounced optical variability is very likely. In the CCD, the SySts seem to fall apart into two subcategories, one group at $\log (f_{\rm X}/f_{\rm opt})$ around $-3$, another group around $-5$. Whether this behavior is related to the $\alpha,\beta,\gamma,\delta$ subtypes and thus could be used as a discriminating and classifying parameter is worth investigating on a larger sample. 

The hardness ratios HR1 and HR2 indicate that all but two systems have reasonably hard spectra, HR1 $>0.3$, which might indicate thermal emission or colliding winds. The two exceptions are StH$\alpha$ 32 (HR1 $=-1.000\pm0.005$) and RR Tel (HR1$=-0.62\pm0.08$), both are known SuperSoft Sources (SSS) and were previously classified as $\alpha$-type SySts. Interestingly, the two SSS SySts both seem to have large distances and the highest X-ray luminosities of $\log L_{\rm X} = 33.7$. RR Tel is a Mira variable and is discussed to have a distance between 2.6\,kpc and 3.5\,kpc \citep{gonzalez-riestra+13}. It has a large negative parallax with small uncertainty, $\pi = -2.89 \pm 0.21$, the implied distance by \cite{bailer-jones+21} is 13\,kpc, in stark contrast to literature values. It is located in the direction of the Galactic bulge, would lie behind the Galactic center, if the \textit{\gai} distance would be taken at face value. At the distance reported in the literature, the X-ray luminosity is about $3\times 10^{32}$\,\lx, still high when compared to the other SySts but within the range spanned by the others. Perhaps more interesting is the case of StH$\alpha$ 32, which has a positive parallax with reasonably small parallax error, $\pi=0.076 \pm 0.016$. \cite{bailer-jones+21} derive a distance between 9 and 12 kpc, which appears reasonable. The object lies at galactic coordinates $l^{II},b^{II} = 197\degr,-30\degr$, and is therefore a halo object as suspected by \cite{luna+13}. With an X-ray luminosity of $L_{\rm X}= 5\times 10^{33}$\,\lx it is the most luminous of all the SySts discussed here but still about two orders of magnitude below other SSS objects.

In Table~\ref{t:sysfrat} the derived X-ray fluxes in the 4XMM catalog \citep{webb+20} and that given for S45 are compared. The spectral model to derive fluxes from count rates is similar for both missions. Both use a power law absorbed by some cold interstellar absorption with a column density of N$_{\rm H} = 3\times 10^{20}$\,cm$^{-2}$. For \xmmn the power law index that was used is $\Gamma =1.7$, for \ero it was $2.0$. Also, the energy bands were not the same originally. The \xmmn fluxes in the band $0.2-2.0$\,keV were scaled to the \ero band ($0.2-2.3$\,keV) with the given spectral model. With the exception of Hen 2-87 and RR Tel all SySts were found to be strongly variable. Hen 3-461 was a factor of $\sim$10 brighter, and RX Pup a factor of $\sim$8 fainter when observed with \ero and compared to \xmmn.

\subsection{The mixed CV samples from Pala, the CRTS, and the SDSS}
Color--magnitude diagrams and CCDs for the three mixed CV samples (Pala, CRTS, SDSS) are shown in Fig.~\ref{f:mixed}. Red and blue symbols have the same meaning as before in identifying objects that were detected in DR1 and both DR1 and S45. None of the CVs has a redder color than $G_{\rm BP}-G_{\rm RP}=2$. All three samples have members that overlap with the main sequence in the CMD. 

The Pala sample is a local one whereas the CRTS and SDSS CVs represent more distant samples. Actually, with a median distance of 974 pc, the CRTS CVs are almost as distant as the secure IPs with a median distance of 1150 pc. Owing to their special detection method they are thought to be dominated by DN. As such, their behavior in the CMD and the CCD and a comparison with the CBcat DN is of interest.  

The CMD and CCD of the CRTS CVs are shown in the middle panel of Fig.~\ref{f:mixed}. Indeed the diagrams seem to indicate a rather homogeneous population. The flux ratio \fxo is confined to a small range, irrespective of the optical color. It is larger by 0.13 dex than for the CBcat DNe. They are also confined in the CMD, in the sense that, for a given optical color, the dispersion in $G_{\rm abs}$ is smaller than for the CBcat DN and for the other standard CVs. The trend is simple: the more luminous the DNe, the redder they appear. 

Both the Pala and the SDSS CV samples appear to be more heterogeneous. Both have very blue and optically luminous members so that they are distributed over 7 to 8 mag in the color range between $G_{\rm BP}-G_{\rm RP}=0$ and 0.2, which also leads to a low flux ratio \fxo for some of their blue members. The median absolute magnitude of the SDSS sample is small and indicates a prevalence of low mass-transfer systems (DN and polars). 

The local sample from Pala also has a rather low median flux ratio \fxo, which is due to two reasons. Their members have a lower median X-ray luminosity than all other standard CVs, and the population misses the optically faint members at $G_{\rm BP}-G_{\rm RP}=0.5$ that other low mass-transfer CVs have. It also has a small value of HR1, probably due to the missing absorption in the vicinity of the sun. 

\begin{table}[t]
\centering
\caption{Median parameters of X-ray nondetected CVs. The upper limit flux was used to calculate the X-ray luminosity and the flux ratio 
$FR = \log (f_{\rm X}/f_{\rm opt})$.}
\label{t:uls}
\begin{tabular}{lrcccc}
\hline\hline
Class & $r_{\rm geo}$ & $\log L_{\rm X}$ & $FR$  & $G_{\rm abs}$ & $B - R$\\
      & (pc) & (\lx) & & (mag) & (mag) \\
\hline
&&&&&\\[-1.5ex]
CBcat DN     & 1075 & 30.85 & $-0.57$ &  9.8 & 0.66 \\
CBcat NL     & 2907 & 31.49 & $-0.96$ &  6.7 & 0.79 \\
Polars       &  800 & 30.45 & $-0.76$ &  9.7 & 0.73 \\
\hline
\end{tabular}
\end{table}

\subsection{Cataclysmic variables not detected as X-ray sources}
In this subsection we briefly discuss the X-ray nondetected CVs in S45. We determined upper limit X-ray fluxes for the standard CVs using the upper limit tool by \cite{tubin+24} for the sensitivity of S45. Using these upper limit fluxes, a diagram with the flux ratios \fxo as a function of the \textit{Gaia} color was generated and is shown in Fig.~\ref{f:uls}. It has entries for the DNe, the NLs, and the polars only. None of the IPs with 4 and 5 stars were missed at the sensitivity of S45. The median properties of the X-ray nondetected CVs are listed in Table~\ref{t:uls}.

The table and the figure show that X-ray nondetection is merely a distance effect. That's due to a combination of the inverse square law (a direct consequence of distance) and the increased interstellar absorption for more distant sources (an indirect consequence of absorption). The median distances of the nondetected subsamples are much larger than those of the detected subsamples, e.g.~800 pc versus 440 pc for the polars. The median luminosity of the nondetected polars is by 0.45 dex smaller, those of the DNe and the NLs by 0.1 dex and 0.75 dex larger than for the detected CVs. This indicates that many objects are below the upper limit flux. The nondetection of polars is also related to their notorious and pronounced accretion-rate and therefore brightness changes. A few of those were in low states during the various \ero all-sky surveys. Indeed, four of the twelve nondetected polars were already discovered by ROSAT, i.e.~could easily be detected by \ero, if they were in their high states. The optical properties of the nondetected members of the three subsamples liste in Table \ref{t:uls} (DN, NL, polars) are rather similar to those of their detected cousins, so that also the flux ratio \fxo is smaller by 0.4 dex for the polars but larger for the DNe and the NLs. One of the nondetected polars, RX J1745.5-2905, was found at a very low flux ratio of only $\log$\fxo $=-2.4$. It is the only object in Fig.~\ref{f:uls} which lies on the branch of the coronal emitters. It is listed with a brightness of G=15.11 (\textit{Gaia} DR3 4057468145836616448) but the ATLAS survey saw it always at around mag 20. We wanted to ascertain its identification as a polar, but could not find a discovery spectrum, it was listed as a polar in the list of CVs by \cite{downes+01}, and therefore the case remained somewhat ambiguous. 

\begin{figure*}
\resizebox{0.49\hsize}{!}{\includegraphics{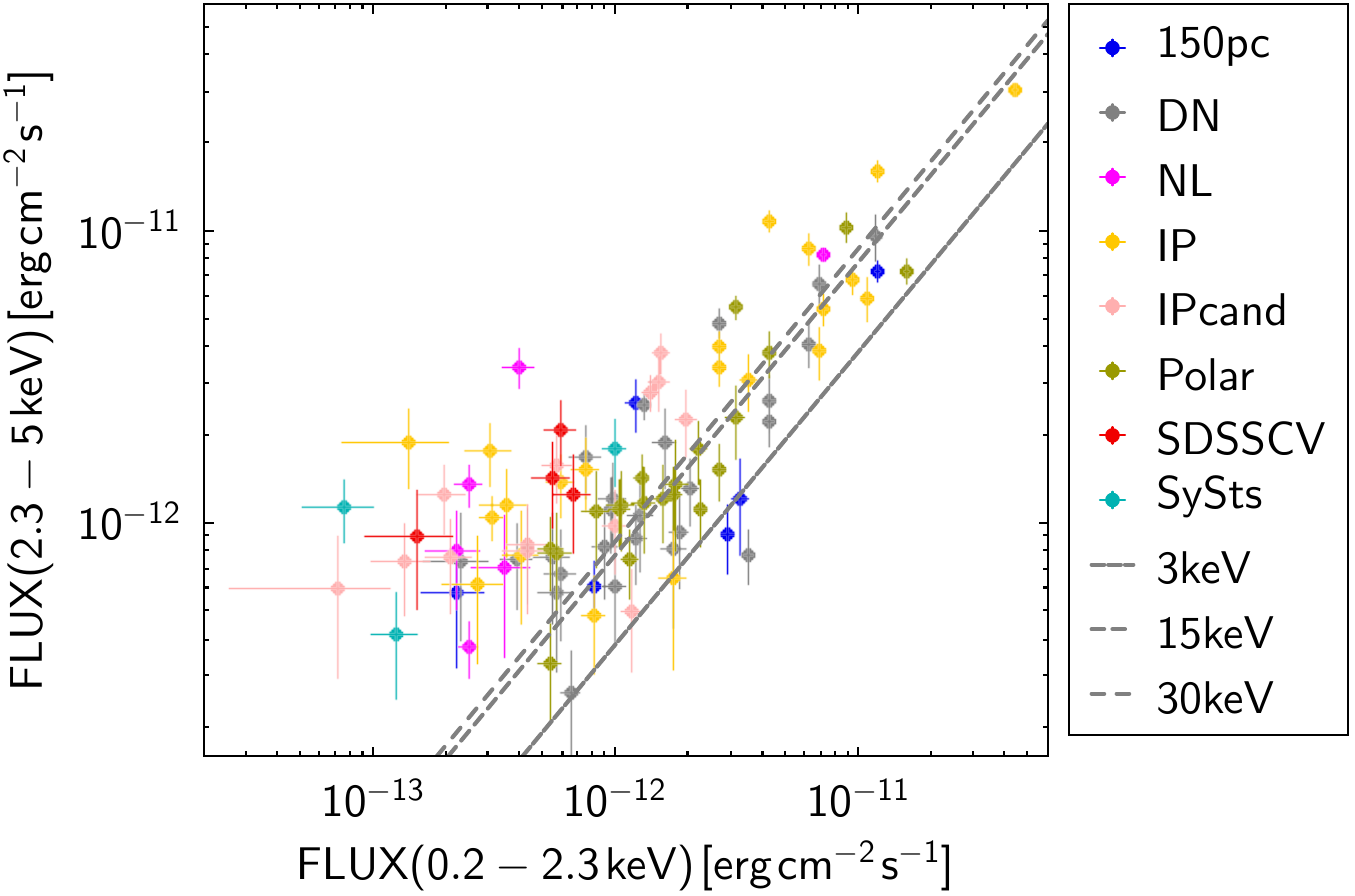}}
\hfill
\resizebox{0.49\hsize}{!}{\includegraphics{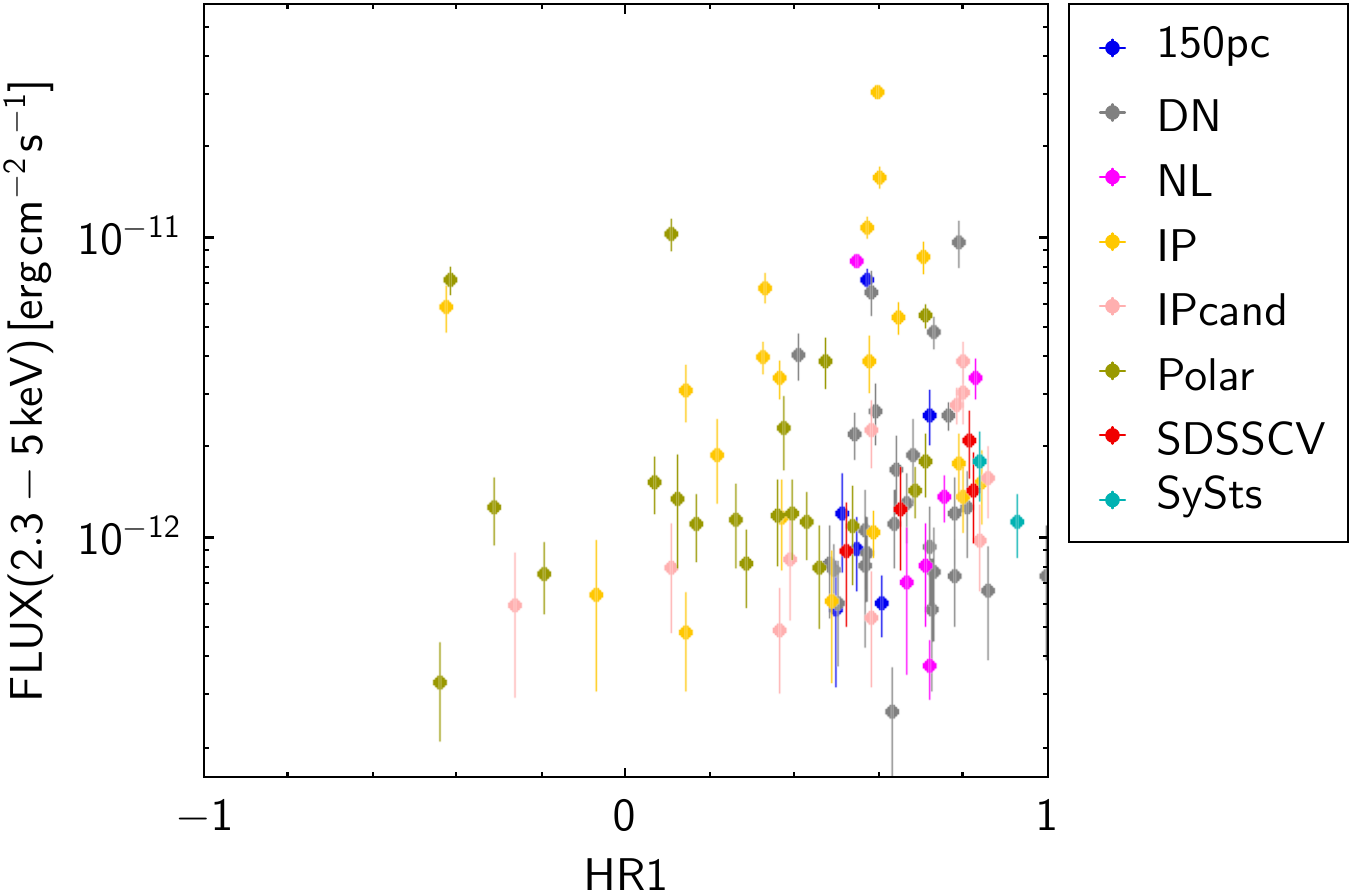}}
\caption{DR1-detected CVs (main catalog) listed also in the hard catalog. The left panel compares the flux in the hard band (2.3-5.0 keV) with the flux in the detection band (0.2-2.3 keV). The diagonal lines illustrate the relation between the fluxes for a thermal plasma of 3, 15, and 30 keV, respectively, from bottom to top. The right panel shows the flux in the hard band as a function of the hardness ratio HR1. The one CRTS CV and the one AM CVn object are not shown to not confuse the diagram.}
\label{f:hard}
\end{figure*}

\begin{figure}
\resizebox{\hsize}{!}{\includegraphics{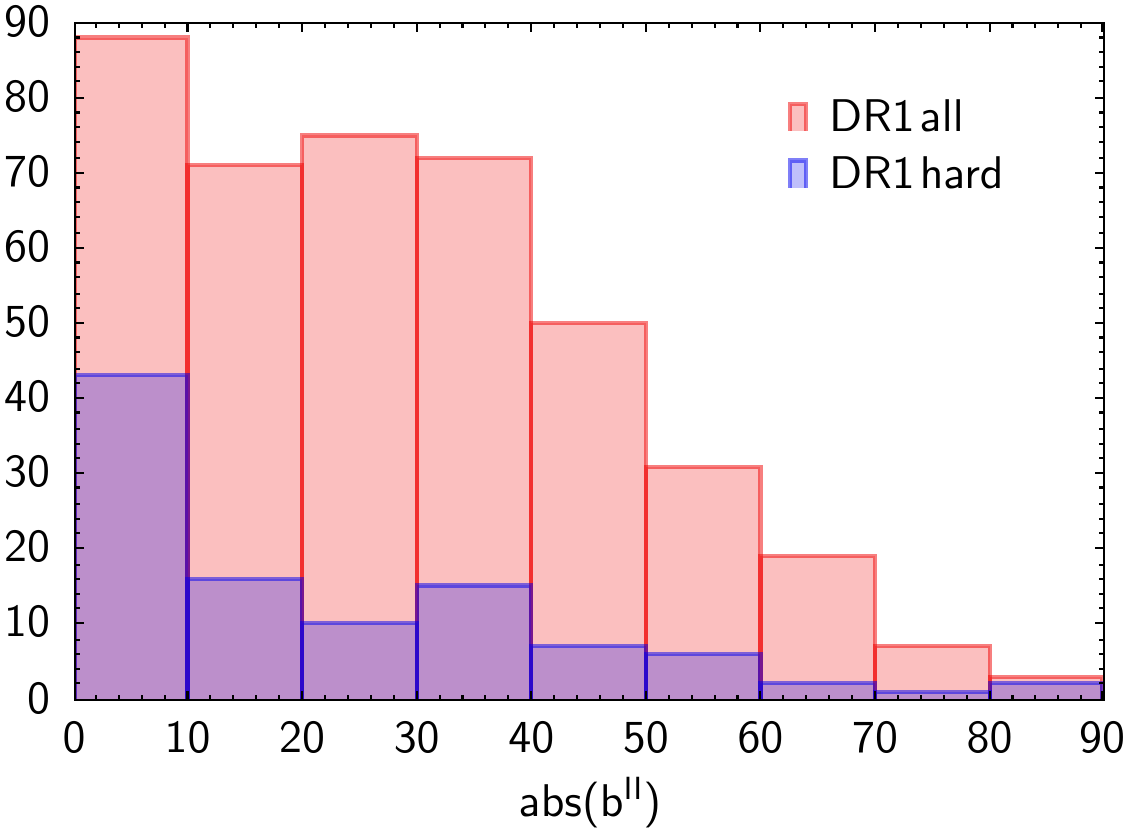}}
\caption{Distribution of all and of hard DR1 CVs as a function of galactic latitude (absolute value).}
\label{f:gal}
\end{figure}

\subsection{Detection summary and hard CVs}
Several CVs are listed in more than one of the catalogs described and characterized above. After the removal of double entries, the list of DR1-detected CVs comprizes 416 objects. This list with some measured and derived quantities is published at the CDS. The structure is illustrated in Table~\ref{t:uniqdr1}. The list of S45-detected CVs comprizes 575 unique systems, which will be published, once the data are fully calibrated.

The main catalog used above, abbreviated as DR1, rests on the single-band detection run using photons between 0.2\,keV and 2.3\,keV. \cite{merloni+24} also published a hard catalog which was based on a 3-band detection run. It comprizes 5466 sources, listing fluxes also in the hard band between 2.3\,keV and 5.0 \,keV. Interestingly, 100 of the 416 DR1 CVs are also listed in the hard catalog, a fraction of a bit less than 2\%  of the sources listed therein (called hard CVs in the following, albeit this might be regarded a misnomer). This fraction is much higher than the fraction of CVs among all DR1 sources. Only two further S45-detected CV systems are also in the hard catalog.

\begin{table}[t]
\centering
\caption{CVs listed in the eRASS1 hard catalog. The fraction given in the third column was computed with respect to all detections in DR1 (column 7 in Table\,\ref{t:matches}).}
\label{t:hard}
\begin{tabular}{lrr}
\hline\hline
    Class & Number & \% DR1 \\
    \hline
&&\\[-1.5ex]
CBcat DN     &  25 & 19\\ 
CBcat NL     &   6 & 26\\
Polars       &  20 & 34\\
IPs (4--5 *) &  21 & 91\\
IPs (2--3 *) &  13 & 57\\
&&\\[-1.5ex]
AM CVns      &   1 &  5\\
SySts        &   3 & 20\\
\hline
&&\\[-1.5ex]
Pala CVs     &   6 & 30\\
CRTS CVs     &   1 &  1\\
SDSS CVs     &   4 &  6\\
\hline
\end{tabular}
\end{table}

The breakdown into object classes is shown in Table~\ref{t:hard} and the flux in the hard band as a function of the flux in the detection band and as a function of the hardness ratio HR1 in Fig.~\ref{f:hard}. Magnetic CVs are predominant among the ahrd CVs. This is due to their high X-ray luminosity and their high-temperatures thermal spectra in the case of the IPs (4 and 5 stars) and the IP candidates (2 and 3 stars), and due to their pronounced thermal spectra and proximity in the case of the polars. DNe and NLs from the CBcat are similarly X-ray luminous to the polars but are slightly more distant, and hence a lower percentage is detected in the hard band. SySts belong to a relatively luminous but distant subpopulation, which explains their small number and fraction of hard CVs. The large average distance of the AM CVns, the CRTS and the SDSS CVs combined with their relatively low luminosity explains their low detection rate as hard CVs. 

The left panel of Fig.~\ref{f:hard} compares the X-ray fluxes in the detection and the hard bands, respectively. The absolute top runner in this diagram, due to its proximity, is EX Hya. At high flux levels, $f_{\rm X} > 2\times10^{-12}$\,\fergs,  many hard CVs follow the flux relation for a nonabsorbed thermal spectrum (the auxiliary lines illustrate the expected flux ratio for thermal plasmas with $kT = 3, 15, 30$\,keV, respectively. Several objects have higher hard fluxes than expected (or lower detection band fluxes than expected), which is indicative of more complex spectra, for example with intrinsic warm absorbers or similar. At low fluxes, $f_{\rm X} (0.2-2.3 \mbox{keV}) < 7 \times 10^{-13}$\,\fergs, all CVs are found above the lines for thermal plasma emission. We interpret this behavior as due to the Malmquist bias (and not due to excessive absorption in the soft band), the hard catalog is reaching its limit at around those fluxes. Among those hard CVs one also finds many soft CVs, those magnetic CVs with an extra soft component in their spectra (right panel of Fig.~\ref{f:hard}). The median hardness ratio HR1 of the hard CVs is essentially the same as that of all DR1 CVs as listed in Table\ref{t:parms}. Taking together, absorption does not seem to be relevant to qualify as a hard CV in eROSITA terms. At the level of detail reached in this current study with all its input biases it is the combination of luminosity, distance, and intrinsic spectral shape that gives rise to its detection in the hard band. 

Figure~\ref{f:gal} gives another interesting aspect into this. Shown there is the distribution of DR1 CVs along the absolute of the galactic latitude, both for all and for the hard CVs. The CVs from the main catalog show a much smoother distribution as function of the galactic latitude compared to the hard CVs. More than 40\% of the hard CVs are found below $b^{II}< 10\degr$, illustrating the large discovery potential of the \ero surveys.

\section{Discussion, conclusion, and outlook}
In this paper, we present an account of the CVs discovered as X-ray sources in the all-sky surveys performed with the \ero instrument on board SRG. We address subsamples that were regarded as representative of typical class members. These are primarily standard CVs with Roche-lobe-filling (mostly) main sequence donors, the nonmagnetic and magnetic DNe and NLs, and the magnetic IPs and polars. We also included close relatives of these objects  in our analysis, namely the DDs and the SySts, which are wind- or RLOF-accreting systems with non main sequence donors. Using the same analysis methods, we also investigated mixed CV samples, the 150 pc sample, and the CRTS and SDSS samples. 

We first sought the main \textit{\gai} parameters for these samples, their brightness, their coordinates, their parallax, and their variability parameters, and then searched for their likely X-ray counterparts from pure positional matching to either published (DR1) or collaboration-internal (S45) \ero catalogs. While in the shallower DR1 catalog the fraction of identified CVs typically lies between 50\% and 60\% (with marked exceptions being the Pala 150 pc sample and the secure IPs reaching more than 90\%), the fraction rises to 80\% or 90\% for the deeper S45. Within a volume with a radius of 500\,pc, the identification fraction is close to 100\% (which is not so for the mixed samples from the CRTS and the SDSS, which reach an identification fraction of 82\% and 88\% in S45 within 500\,pc). The given numbers are similar for the DDs, but not so for the SySts. For both these nonstandard CVs, new X-ray detections are reported here, but the identification fraction of distant SySts at the limit of S45 remains as low as 13\%. 

We used the combined \ero X-ray and \textit{\gai} optical parameters to derive a few typical average parameters of the studied samples and to locate the individual objects in a variety of diagnostic diagrams. These combine the distance, luminosity, absolute magnitude, optical and X-ray colors, optical variability amplitude, and the ratio between X-ray and optical fluxes. 

We used the X-ray hardness to separate (magnetic) CVs with purely thermal spectra from those that have a soft component, and performed a spectral analysis of the brightest polars and likely soft IPs. We were therefore able to measure blackbody temperatures for a number of bright polars (in the range of 20 - 60 eV) to quantify the flux ratio between the hard X-ray emission from the cooling accretion plasma in the column and the soft X-ray emission of reprocessed origin as a function of the magnetic field strength. This ratio shows a trend toward low values with higher $B$ by about a factor of $\sim$100, albeit less pronounced than the trend seen in a \ros analysis based on an assumed constant blackbody temperature of 25 eV. We were also able to measure blackbody temperatures for seven soft IPs, finding values of between 47 eV and 96 eV. The temperature of the four known soft IPs agrees well with an earlier analysis. Three soft IPs are reported here for the first time with temperatures of between 60 and 95 eV. None of the soft IPs seem to be dominated by the soft X-ray emission.

Seven of the 20 detected SySts are reported as X-ray sources here for the first time. A comparison of the X-ray fluxes of SySts that were previously observed with \xmmn shows variations of up to a factor of 10. The identified fraction of SySts is the lowest of all subclasses and of all the samples studied here: just 13\% in the stack of all \ero surveys. A much deeper survey is needed to detect the SySts at X-ray wavelengths in significant numbers.

We also analyzed the X-ray nondetected CVs on the basis of their upper-limit X-ray flux. This suggests that their nondetection is mainly a distance effect and that accretion rate changes (low states) are only contributing for the polars.

In order to estimate the fraction of chance associations, we checked the CVs that are also considered counterparts to eRASS:1 (Salvato et al., in prep). These latter authors used NWAY \citep{salvato+18} to identify the most likely counterparts to eRASS1 X-ray sources, considering the separation, positional errors, and density of sources in \textit{\gai} DR3 and eRASS1, respectively, together with a prior based on the \textit{\gai} properties of sample of 24000 X-ray sources from 4XMM used as a training sample. The agreement between the two approaches is very high (95\%); only for 21 objects do the two approaches select a different optical counterpart. For 15 sources in disagreement, the counterparts have a very low probability ($p\_any < 0.05$) of being correct in Salvato et al.~(2024). However, for 6 sources ($\sim$1\%), the NWAY algorithm points to another counterpart with high probability ($p\_any>0.9$). These six objects are listed in Table~\ref{t:nway}. No systematic errors were recovered in regard to the nature or properties of those CVs that might lead to possible confusion with another source. The six sources were checked individually, and also with the help of the SIMBAD database and optical finding charts. As a result, it appears that just one of the identifications proposed here appears to be unlikely (1eRASS J131559.7-370018 identified with the SySts object CD-36 8436). In two further cases, the identification with a CV appeared most likely, but it is also possible that a nearby object, such as a background AGN, could contribute to the \ero-detected X-ray source. In one case ---where a nearby AGN (separation of only $3.9"$) was spectroscopically identified--- it was shown that it does not contribute to the X-ray signal \citep{vogel+08}. The identification rate of more than 95\% of CVs with NWAY is a good demonstration of its reliability in finding correct counterparts. Some improvements can be expected if specially adapted training samples are used. Currently, CVs and related objects are under-represented compared to the many AGNs and stars in the training samples. This is simply due to the fact that, despite decade-long CV research, the sample sizes and our knowledge about the combined X-ray and optical properties of the various classes are still limited. Hence, without specially designed counterpart-finding algorithms, the more exotic objects, such as CVs and related systems, are likely to be overlooked.

The volumetric 150 pc sample has the lowest X-ray luminosity on average. If this sample represents an intrinsic population, then all other samples are biased towards a higher $L_{\rm X}$. However, this sample is small, and only 22 of the original 42 CVs can be studied with publically available \ero data. Prospects to significantly increase the volumetric sample with a larger radius are excellent. One can discover objects at the median luminosity of the 150 pc sample ($\log (L_{\rm X}/\mbox{erg\,s}^{-1})=30.34$) out to 600 pc. The faintest objects in this sample at $\log (L_{\rm X}/\mbox{erg\,s}^{-1}) =29.3$ can be detected out to a radius of 300\,pc.

Finding those among the million or more other X-ray sources detected by \ero (mostly AGNs and stellar coronal emitters) requires careful selection of the likely optical counterparts and comprehensive spectroscopic follow up. The combined X-ray and optical properties of CVs as presented in this paper were used to select likely CWDBs from the various \ero surveys and a rigorous identification program has been initiated. The results of the optical follow-up program will be reported in forthcoming papers.

\begin{acknowledgements}
We gratefully acknowledge constructive criticism of our referee. \\

This work was supported by the German DLR under contract 50 OR 2203 and the Deutsche Forschungsgemeinschaft under grant numbers Schw536/37-1 and Schw536/38-1.\\


This work is based on data from \ero, the soft instrument aboard SRG, a joint Russian-German science mission supported by the Russian Space Agency (Roskosmos), in the interests of the Russian Academy of Sciences represented by its Space Research Institute (IKI), and the Deutsches Zentrum für Luft- und Raumfahrt (DLR). The SRG spacecraft was built by Lavochkin Association (NPOL) and its subcontractors, and is operated by NPOL with support from the Max Planck Institute for Extraterrestrial Physics (MPE). \\

The development and construction of the eROSITA X-ray instrument was led by MPE, with contributions from the Dr. Karl Remeis Observatory Bamberg \& ECAP (FAU Erlangen-Nuernberg), the University of Hamburg Observatory, the Leibniz Institute for Astrophysics Potsdam (AIP), and the Institute for Astronomy and Astrophysics of the University of Tübingen, with the support of DLR and the Max Planck Society. The Argelander Institute for Astronomy of the University of Bonn and the Ludwig Maximilians Universität Munich also participated in the science preparation for eROSITA. \\

The eROSITA data shown here were processed using the eSASS/NRTA software system developed by the German eROSITA consortium. \\

\end{acknowledgements}

\bibliographystyle{aa}
\bibliography{cvsdr1}

\begin{appendix}
\section{Known CVs in DR1}
~~~~~
\begin{table}[h!]
\centering
\caption{Known CVs detected in \ero DR1 (extract). The full list of 416 CVs is published at the CDS. Given are per object the IAU name from DR1, the CV name and its type (IPs with 4 and 5 stars are labeled IPs, those with 2 and 3 stars are labeled IPcands), the \textit{Gaia} source ID, the X-ray coordinates (RA and DEC), the separation between the X-ray and the optical (\textit{\gai}) coordinates in arcsec, the number of photons collected in DR1 in the energy band $0.2-2.3$ keV, the logarithm of the derived flux (in erg cm$^{-2}$ s$^{-1}$), the \textit{\gai} $G$-band magnitude, the distance (in pc), the logarithm of the derived X-ray luminosity (in erg s$^{-1}$), the X-ray to optical flux ratio, the absolute magnitude in the $G$-band, and the \textit{\gai} color.}
\label{t:uniqdr1}
\begin{tabular}{rlllr}
\hline\hline
  \multicolumn{1}{c}{Seq} &
  \multicolumn{1}{l}{IAU NAME} &
  \multicolumn{1}{l}{CV Name} &
  \multicolumn{1}{l}{CV type} &
  \multicolumn{1}{l}{GaiaID} \\
\hline
  1 & 1eRASS J000049.6-771856 & BE Oct & CBcat DN & 4684042885786567936\\
  2 & 1eRASS J001449.4-523226 & CRTS\_J001449.5-523215 & CRTS CV & 4972555149231837952\\
  3 & 1eRASS J002130.9-571919 & J0021-5719 & CBcat DN & 4918835764173746432\\
  4 & 1eRASS J002430.2-663544 & J0024-6635 & CBcat DN & 4707183143081508480\\
  5 & 1eRASS J004214.2-560920 & J0042-5609 & CBcat DN & 4920350474584208384\\
  6 & 1eRASS J012852.4-233947 & EQ Cet & Polar & 5041907811522399488\\
  7 & 1eRASS J013330.6-810751 & Gaia23bpl & Polar & 4630500071128655360\\
  8 & 1eRASS J014100.5-675326 & BL Hyi & Polar & 4697621824327141248\\
  9 & 1eRASS J015400.9-594747 & RX J0153-59 & Polar & 4714563374364671872\\
  10 & 1eRASS J020950.8-631839 & WX Hyi & CBcat DN & 4701058897674276736\\
  11 & 1eRASS J021440.3-265605 & CRTS\_J021440.7-265602 & CRTS CV & 5117799471325639424\\
  12 & 1eRASS J021927.6-304545 & AX For & CBcat DN & 5067753236787919232\\
  13 & 1eRASS J022046.8-144826 & CRTS\_J022047.3-144823 & CRTS CV & 5146608393960138752\\
  14 & 1eRASS J022137.6-261953 & CRTS\_J022137.8-261952 & CRTS CV & 5117190960359055360\\
  15 & 1eRASS J023051.3-684206 & CW Hyi & Polar & 4695093944014478080\\
  16 & 1eRASS J023611.4-521915 & WW Hor & Polar & 4744804750196738560\\
  17 & 1eRASS J024353.0-160307 & J0243-1603 & CBcat DN & 5132953176802158976\\
  18 & 1eRASS J024607.4-633548 & RU Hor & CBcat DN & 4721232870163826816\\
  19 & 1eRASS J025248.3-395910 & J0252-3959 & CBcat DN & 4949102295630055552\\
  20 & 1eRASS J025533.2-475042 & ASASSN-14ei & AM CVn & 4752105374261391360\\
\hline
\end{tabular}
\vspace{1cm}

\begin{tabular}{rrrrrrlrlrrr}
\hline\hline
\multicolumn{1}{c}{Seq} &
  \multicolumn{1}{c}{RA} &
  \multicolumn{1}{c}{DEC} &
  \multicolumn{1}{c}{Sep} &
  \multicolumn{1}{c}{CTS} &
  \multicolumn{1}{c}{$\log F_{\rm X}$}&
  \multicolumn{1}{c}{$G$} &
  \multicolumn{1}{c}{Dist} &
  \multicolumn{1}{c}{$\log L_{\rm X}$} &
  \multicolumn{1}{c}{$\log F_{\rm X}/ F_{\rm opt}$} &
  \multicolumn{1}{c}{$G_{\rm abs}$} &
  \multicolumn{1}{c}{$B-R$} \\
\hline
  1 & 0.20702 & -77.31580 &  2.6 &  17 & -13.03 & 18.83 & 1472 & 31.39 & -0.64 &  7.99 & 0.52\\
  2 & 3.70616 & -52.54079 & 11.2 &   6 & -13.37 & 20.03 &  974 & 30.68 & -0.50 & 10.09 & 0.60\\
  3 & 5.37882 & -57.32222 &  2.2 & 149 & -12.01 & 17.81 &  611 & 31.64 & -0.02 &  8.88 & 0.67\\
  4 & 6.12621 & -66.59569 &  8.7 &   7 & -13.35 & 19.75 & 1185 & 30.87 & -0.60 &  9.38 & 0.63\\
  5 & 10.55951 & -56.15567 & 0.5 & 141 & -12.04 & 17.72 &  635 & 31.64 & -0.09 &  8.71 & 0.63\\
  6 & 22.21860 & -23.66326 & 3.7 &  79 & -12.23 & 17.57 &  283 & 30.75 & -0.35 & 10.31 & 0.72\\
  7 & 23.37767 & -81.13084 & 1.8 &  51 & -12.64 & 18.37 &  429 & 30.70 & -0.43 & 10.20 & 0.99\\
  8 & 25.25214 & -67.89066 & 0.8 & 571 & -11.58 & 17.21 &  129 & 30.72 &  0.17 & 11.66 & 0.79\\
  9 & 28.50403 & -59.79666 & 1.3 & 314 & -11.76 & 15.76 &  321 & 31.33 & -0.60 &  8.23 & 0.36\\
  10 & 32.46193 & -63.31087 & 0.8& 874 & -11.37 & 14.65 &  228 & 31.43 & -0.65 &  7.85 & 0.63\\
  11 & 33.66799 & -26.93487 & 6.0 & 12 & -13.17 & 20.22 & 1536 & 31.28 & -0.22 &  9.28 & 0.53\\
  12 & 34.86524 & -30.76264 & 4.4 &112 & -12.24 & 17.87 &  349 & 30.93 & -0.23 & 10.16 & 0.69\\
  13 & 35.19538 & -14.80737 & 6.4 & 13 & -13.12 & 19.20 & 2311 & 31.69 & -0.57 &  7.38 & 0.99\\
  14 & 35.40700 & -26.33159 & 2.9 & 37 & -12.71 & 19.64 & 1025 & 31.39 &  0.01 &  9.59 & 0.37\\
  15 & 37.71381 & -68.70194 & 1.7 &316 & -11.94 & 17.53 &  543 & 31.60 & -0.07 &  8.86 & 0.87\\
  16 & 39.04773 & -52.32092 & 1.4 & 89 & -12.38 & 19.35 &  334 & 30.75 &  0.22 & 11.73 & 1.16\\
  17 & 40.97118 & -16.05213 & 15.5 & 5 & -13.59 & 20.26 &  507 & 29.90 & -0.63 & 11.74 & 0.12\\
  18 & 41.53112 & -63.59692 & 2.7 & 41 & -12.74 & 19.84 &  582 & 30.87 &  0.06 & 11.01 & 0.73\\
  19 & 43.20127 & -39.98624 & 1.6 & 44 & -12.68 & 19.05 &  405 & 30.61 & -0.20 & 11.01 & 0.56\\
  20 & 43.88855 & -47.84509 & 0.3 &177 & -12.13 & 16.46 &  257 & 30.77 & -0.68 &  9.42 & 0.54\\
\hline
\end{tabular}

\end{table}

\begin{landscape}

\begin{table}
\small
\caption{Six objects with different counterpart identification presented in this work and with NWAY (Salvato et al., in prep.). Given are the IAU name, the CV name, its type, the \gai DR3 ID of the counterpart chosen in this work, the $G$-band magnitude, the implied distance and the positional offset between  the \gai and the 1eRASS source, the \gai DR3 ID of the counterpart chosen by NWAY, its $G$-band magnitude, parallax, the positional offset, and the final likely identification.}
\label{t:nway}
{\centering    
\begin{tabular}{lllrrrrrrrrr}
  \multicolumn{1}{l}{IAUNAME} &
  \multicolumn{1}{l}{CVName} &
  \multicolumn{1}{l}{Type} &
  \multicolumn{1}{c}{GaiaID\_CV} &
  \multicolumn{1}{c}{G} &
  \multicolumn{1}{c}{Dist.} &
  \multicolumn{1}{c}{$\Delta x$} &
  \multicolumn{1}{c}{GaiaID\_NWAY} &
  \multicolumn{1}{c}{G} &
  \multicolumn{1}{c}{$\pi$} &
  \multicolumn{1}{c}{$\Delta x$ } &
  Counterpart\\
 & & & & (mag) & (pc) & ($"$) & & (mag) & (mas) & ($"$) \\ 
 \hline
1eRASS J054320.2-410154 & TX Col & IP & 4804695427734393472 & 15.61 & 909 & 2.0 & 4804695423438691200 & 15.13 & 1.02 & 0.6 & CV\\
1eRASS J061145.8-814921 & AH Men & CBcat NL & 5207385651533430912 & 13.78 & 491 & 4.4 & 5207384891323130368 & 13.48 & 2.04 & 2.6 & CV\\
1eRASS J101545.8+033318 & J1015+0333 & CBcat DN & 3859948269249864832 & 17.76 & 1641 & 7.4 & 3859948273545278336 & 20.4 & --- & 6.5& CV (+ ?)\\
1eRASS J131223.5+173658 & 2XMM J1312 & Polar & 3937217307886601088 & 19.53 & 830 & 1.8 & 3937217307885385728 & 19.5 & $-0.2$ &3.9 & CV\\
1eRASS J131559.7-370018 & CD-36 8436 & SySts & 6165818495673620352 & 9.8 & 2260 & 21.1 & 6165817018204434816 & 11.4 & 3.12 & 2.4 & star (?)\\
1eRASS J132558.4-145223 & ZTF19abdsnjm & AM CVn & 3608015074033774336 & 20.4 & 1160 & 5.9 & 3608015074033776768 & 18.4 & 0.0 & 8.4 & CV (+ ?)\\
\hline
\end{tabular}
}

Notes for individual objects: \\
1) TX Col, an IP, which are notorious and bright X-ray emitters. X-ray source in HEAO, Swift, 1RXS. Secure CV identification \hfill \\
2) AH Men: this NL CV is listed as H, RXS, Swift source. Secure CV identification\\
3) J1015+0333: the NWAY ctp is faint, at same positional offset as the CV, not labeled as extra-galactic. The implied high $f_{\rm X}/f_{\rm opt}$ excludes a coronal emitter as  X-ray source. Implied optical/X-ray properties typical of a DN. Contribution by background AGN not completely excluded but unlikely. \\
4) 2XMM J1312: \cite{vogel+08} identified the eclipsing polar as counterpart to the X-ray source. They also presented a spectrum of the AGN chosen by NWAY as the likely counterpart, but the periodic modulation of the X-ray signal by 100\% excludes any significant contribution of the AGN to the measured X-ray signal. \\
5) CD-36 8436: It is an optically bright SySts, the positional offset of our proposed optical counterpart is rather large. The X-ray position is confirmed from the living \swi catalog \footnote{\url{https://www.swift.ac.uk/LSXPS/}}. Both NWAY and the HamStars code \citep{freund+24} select a positionally better matching \gai object as likely counterpart, a stellar coronal emitter. A possible or even likely CV mismatch.\\
6) ZTF19abdsnjm: The chosen counterpart by NWAY is brighter (but still implying a high X-ray to optical flux ratio), at larger positional offset, and flagged as extragalactic. While the implied the optical and X-ray properties are reasonable for a CV identification, a contribution from a background AGN cannot be excluded. 
\end{table}

\end{landscape}

\end{appendix}
\end{document}